\documentstyle[multicol,aps,prl,eqsecnum,epsf]{revtex}
\renewcommand{\narrowtext}{\begin{multicols}{2} \global\columnwidth20.5pc}
\renewcommand{\widetext}{\end{multicols} \global\columnwidth42.5pc}  
\def\top#1{\vskip #1\begin{picture}(290,80)(80,500)\thinlines \put(   
65,500){\line( 1, 0){255}}\put(320,500){\line( 0, 1){  
5}}\end{picture}}
\def\bottom#1{\vskip #1\begin{picture}(290,80)(80,500)\thinlines \put(
330,500){\line( 1, 0){255}}\put(330,500){\line( 0, -1){
5}}\end{picture}}

\begin{document}
\title{Lowest-Landau-level theory of the quantum Hall effect:\\
the Fermi-liquid-like state}
\author{N. Read}
\address{Departments of Physics and Applied Physics, 
Yale University,\\ P.O. Box 208120, New Haven, Connecticut 06520}
\date{\today}
\maketitle
\newcommand{\be}{\begin{equation}}
\newcommand{\ee}{\end{equation}}
\newcommand{\bea}{\begin{eqnarray}}
\newcommand{\eea}{\end{eqnarray}}
\newcommand{\non}{\nonumber}
\newcommand{\br}{{\bf r}}
\newcommand{\ba}{{\bf a}}
\newcommand{\bA}{{\bf A}}
\newcommand{\bk}{{\bf k}}
\newcommand{\bq}{{\bf q}}
\newcommand{\pt}{\tilde{\phi}}
\newcommand{\zb}{\bar{z}}
\newcommand{\wb}{\overline{w}}
\newcommand{\bv}{{\bf v}}
\newcommand{\bg}{{\bf g}}
\newcommand{\bj}{{\bf j}}
\begin{abstract}
A theory for a Fermi-liquid-like state in a system of charged bosons at
filling factor one is developed, working in the lowest Landau level. The
approach is based on a representation of the problem as fermions with a
system of constraints, introduced by Pasquier and Haldane (unpublished). 
This makes the system a gauge theory with gauge algebra $W_\infty$. The
low-energy theory is analyzed based on Hartree-Fock and a corresponding
conserving approximation. This is shown to be equivalent to introducing a
gauge field, which at long wavelengths gives an infinite-coupling U(1)
gauge theory, {\em without} a Chern-Simons term. The system is
compressible, and the Fermi-liquid properties are similar, but not
identical, to those in the previous U(1) Chern-Simons fermion theory. 
The fermions in the theory are effectively neutral but carry a dipole
moment. The density-density response, longitudinal conductivity, and the
current density are considered explicitly.

 \end{abstract}
\narrowtext



\section{Introduction}
\label{intro}

The so-called composite-particle view of the liquid states of electrons
(or other charged particles) in two-dimensions in a high magnetic field
\cite{book} has been developed gradually over more than a decade
\cite{girvin,gm,laughan,zhk,jain,lopfrad,fishlee,read87,read89,hlr,read94}.
Girvin \cite{girvin} proposed to develop a Ginzburg-Landau theory of the
fractional quantum Hall effect, with an action for a complex
scalar (boson) field and containing a Chern-Simons
(CS) term to enforce the condition that the quantized vortices carry
fractional charge. Girvin and MacDonald \cite{gm} introduced a singular
gauge transformation and exhibited algebraic long-range order in a bosonic
field. This transformation, which attaches $\delta$-function flux tubes to
particles (via a CS term in the action of the field theory) and so in
general changes the statistics of the particles as in
the theory of anyons \cite{wilczek}, was then used in
several theories, in conjunction with the mean-field approximation of
replacing the gauge field strength by its expectation value, to obtain
a system in a different magnetic field. Thus, anyon superconductivity was
discovered by mapping anyons in zero magnetic field to fermions filling
Landau levels in a magnetic field \cite{laughan}; the Laughlin states
\cite{laugh} were described by mapping fermions to bosons
in zero net magnetic field and then Bose-condensing them \cite{zhk}; the
Laughlin and hierarchy \cite{hald83,halp84} states were reinterpeted by
mapping fermions to fermions in a reduced magnetic field and then filling
Landau levels \cite{jain,lopfrad}; the hierarchy states and the anyon
superconductors in zero magnetic field were redescribed by hierarchical
extension of the mapping to bosons, using duality methods \cite{fishlee}. 
At the same time, a lowest Landau level (LLL) treatment of the
Ginzburg-Landau idea was developed \cite{read87,read89}, without using
$\delta$-function flux tubes, by attaching vortices to electrons to
convert them to bosons; in this case, the bosons condense and have true
long-range order. 

It has to be admitted that these ways of viewing the fractional quantum
Hall effect produced little in the way of distinctive experimental
predictions or explanations that were not already known by other methods,
though interesting speculations concerning the phase transitions between
the quantized Hall plateaus \cite{klz} may be an exception. The situation
changed, however, following the discovery of an anomaly in the surface
acoustic wave propagation at filling factor $\nu=1/2$ (and less strongly
at other filling factors, such as $1/4$, $3/2$) \cite{willett90}. This
result speeded the development of a theory \cite{hlr} (to be referred
to as HLR) for a case not
included in the above list, in which fermions (electrons) are mapped to
fermions at zero magnetic field and form a Fermi sea. In the simplest
cases, this occurs for filling factor $\nu=1/2$, $1/4$, $1/6$, \ldots. 
The Fermi sea was predicted to be a compressible state that does not
produce a Hall plateau, and the experimental result of a longitudinal
conductivity increasing linearly with wavevector \cite{willett93a} was
explained \cite{hlr}. The Fermi surface, at which the fermions
exist as genuine low-energy excitations, was observed through 
geometric resonance effects at $\nu$ close to $1/2$ in further 
surface acoustic wave experiments \cite{willett93b} (as predicted
explicitly in Ref.\ \cite{hlr}), and in other experiments 
\cite{stormerferm,goldman}. (We should point out that for other filling
factors in the fermion description, the fermions are dressed to become the
fractionally-charged, fractional-statistics quasiparticles
\cite{laugh,hald83,halp84,asw}, so are not observed as fermions.)  

In this paper, we return to the basic theory of the Fermi-liquid-like
state. Recent work \cite{sm,dhlee,ph} has raised the possibility of
changes in the way we think about the theory of the low energy excitations
near the Fermi surface. In particular, these authors find contraints not
mentioned in any earlier papers known to the present author. At the same
time, we may be motivated by trying to avoid the seemingly artificial CS
approach, which begins with a singular gauge transformation. Ultimately, 
it would aid our understanding to have more intuition about
what drives the formation of the Fermi-liquid-like (and other) states. 
There are no flux tubes attached to the particles in reality; the
background magnetic field remains essentially uniform in these states of
matter. The approach begun in Ref.\ \cite{read89} was intended to head in
this direction. It uses LLL states only, so is valid in the (not entirely
realistic) limit of interactions weak compared with $\omega_c$, and binds
{\em vortices} to the electrons to lower the energy, thus forming the
composite particles. Several implications of this approach were pointed
out in Ref.\ \cite{read94} for the Fermi sea and the Bose condensate. 

The approach taken in the present paper avoids the CS approach. While it is
perhaps not as simple-minded as one would want, it does make close
contact with the work just cited \cite{read94}. Here we start {}from an
approach of Pasquier and Haldane (PH) \cite{ph,haldpriv}, that gives an
exact representation of the LLL problem in the case of charged bosons in a
magnetic field at $\nu=1$, where a Fermi liquid (FL) state is possible.
Although our paper is long and fairly detailed, we can give a succinct
summary of our results. {\em The low-energy, long-wavelength theory is a
FL coupled to a gauge field} (not to be confused with the physical
electromagnetic field). In contrast to the scenario arising \cite{hlr} in 
the CS (singular gauge transformation) approach, {\em there is no CS term
in this low energy theory}. Consequently, the gauge field is said to be
``strongly-coupled'' and one of its effects is to enforce constraints that
agree with those of \cite{sm,dhlee,ph}. This in turn has the effect of
making the fermions uncharged, but they are left with a subleading
coupling to electromagnetic fields through a dipole moment. The interplay
of this moment with the transverse part of the gauge field leads to a
finite compressibility, in spite of the neutrality of the particles. It
also leads to the CS equations, that relate the curl of the vector
potential to the density, and a similar equation for the current, still
being valid, in spite of the absence of a CS term in the action, in
agreement with Ref.\ \cite{read94}. In general, the good agreement with
experimentally-observed phenomena achieved in the theory of HLR is not
spoilt in the present theory. Nonetheless, the detailed structure of this
FL-like theory is modified. While the theory is developed here for
$\nu=1$ bosons, there are many indications that the results are more
general. These include the derivation in Ref.\ \cite{sm} for general
number of attached flux.

Sec.\ \ref{review} contains a more detailed review of previous work, and a
more detailed overview of the paper. In Sec.\ \ref{PHapproach}, we explain
the formalism due to Pasquier and Haldane that will be used in this paper. 
In Sec.\ \ref{HFconserving}, we perform explicit calculations of response
functions, including those for the constraint operators, and interpret the
results in terms of a strongly-coupled gauge field. In Sec.\
\ref{allorders} we outline the extension of the results to all orders, and
provide some general discussion. Sec.\ \ref{conclusion} is the conclusion.
Appendix \ref{noncom} discusses some details of the formalism, including
the noncommutative Fourier transform, and Appendix \ref{hubb} indicates
how a Hubbard-Stratonovich transformation can be used.


\section{Review and overview}
\label{review}
In this Section, we review some of the background necessary for the
discussion in this paper. We begin with the U(1) Chern-Simons (CS) fermion
approach developed in Ref.\ \cite{hlr}. The
Fermi liquid-like state proposed in that paper is the main topic of the
present work; however, we will not review the relation to experiments. 
In Subsec.\ \ref{physical} we review ``physical'' pictures which are
based on consideration of the wavefunctions of the system, as opposed to
field theoretic methods. In Subsec.\ \ref{recent}, we review recent work
which attempts to push the U(1) CS approach down to a low-energy effective
theory in the lowest Landau level (LLL). Finally, in Subsec.\
\ref{overview}, we give a brief overview of the main results and of the
layout of the remainder of the paper.

\subsection{U($1$) Chern-Simons fermion theory}
\label{chern}

In this approach the particles are represented as fermions with a
$\delta$-function of flux attached, of strength an integral
number $\tilde{\phi}$ of flux quanta $\Phi_0$. Then the underlying
particles must be bosons when $\tilde{\phi}$ is an odd integer,  and
fermions when $\tilde{\phi}$ is even (for noninteger $\tilde{\phi}$, the
underlying particles must be anyons). We will reserve the term
``particles'' for these original particles, and refer to the
transformed particles as ``fermions'' or ``quasiparticles''. The imaginary
time action
(see, e.g.\ \cite{hlr}, to be referred to as HLR) is (in the gauge where
$\nabla\cdot{\bf a}=0$)
\bea
S&=&\int d\tau\, d^2r\left[\psi^\dagger\left(\frac{\partial}{\partial\tau}
-ia_0-\mu\right)\psi\right.\non\\
&&\left.\mbox{}+\frac{1}{2m}\left|(-i\nabla-{\bf a}-{\bf
A})\psi\right|^2-\frac{i}{2\pi\tilde{\phi}}a_0\nabla\wedge{\bf
a}\right]\non\\
& &\mbox{}+\frac{1}{2}\int d\tau\,d^2r\,d^2r'\,V({\bf r}-{\bf
r}')\psi^\dagger({\bf r})
\psi^\dagger({\bf r}')\psi({\bf r}')\psi({\bf r}).\non\\
&&
\eea
Here $\psi$ is the field operator for the fermions, rather than that for
the underlying particles, which could be fermions (electrons) or bosons.
We will use the notation (note the use of the summation convention for
repeated Greek indices)
\begin{equation}
{\bf a}\wedge{\bf b}=\varepsilon_{\mu\nu}a_\mu b_\nu
\end{equation}
for a cross product of vectors $\bf a$, $\bf b$ in two dimensions, 
$\mu$, $\nu,\ldots=x$, $y$ to label the two components, and
$\varepsilon_{\mu\nu}=-\varepsilon_{\nu\mu}$, $\varepsilon_{xy}=1$ for the
two-dimensional alternating tensor. We have set $\hbar=1$ and, starting
with gaussian units, we have absorbed
$-e$ into the scalar potential and electric field, and $(-e/c)$
into the vector potential and magnetic field, so the charge of the
particles is one and the flux quantum is $2\pi$. The uniform background
magnetic field is $\nabla\wedge{\bf A}=B>0$ which corresponds to 
the negative $\hat{z}$ direction (in the three-dimensional sense) in
conventional units. We choose the unit of length so that the magnetic
length $\ell_B^{-2}=B=1$. It will also be convenient to write
$\wedge{\bf a}$ for the vector whose components are $(\wedge{\bf a})_\mu=
\varepsilon_{\mu\nu}a_\nu$; then ${\bf a}\cdot\wedge{\bf b}={\bf
a}\wedge{\bf b}$.

Varying $a_0$ in the action leads to 
\begin{equation}
\nabla\wedge{\bf a}=-2\pi\tilde{\phi}\rho,
\end{equation}
where $\rho({\bf r})=\psi^\dagger({\bf r})\psi({\bf r})$ is the number
density both of the Chern-Simons fermions and of the underlying particles.
When the filling factor $\nu=2\pi\bar{\rho}/B$ is $1/\tilde{\phi}$
(where $\bar{\rho}$ is the average density),
there is no net field for the fermions, and within a mean-field
approximation, a Fermi sea ground state is possible.

The leading approximation for the linear-response functions is the random
phase approximation (RPA), in both the gauge field $a_0$, ${\bf a}$ and
the Coulomb (or other) interaction $V({\bf r})$. In Fourier space the full
density-density response function is then \cite{hlr}, before any
approximation,
\begin{equation}
\chi_{\rho\rho}=\frac{\chi^{\rm irr}_{\rho\rho}}{1+V({\bf q})\chi^{\rm
irr}_{\rho\rho}},
\end{equation}
and in the RPA $\chi^{\rm irr}_{\rho\rho}=\chi_0^{\rm irr}$, where
\begin{equation}
\chi_0^{\rm irr}=\frac{\chi_0}{1-(2\pi\tilde{\phi})^2\chi_0\chi_0^\perp/q^2}.
\label{densrespinchern}
\end{equation}
Here $\chi^{\rm irr}$ is the response function which is irreducible with
respect to the interaction $V$ {\em only} (i.e., diagrammatically, it does
not become disconnected when a single interaction line is cut), while
$\chi_0$ is the density-density response for the non-interacting sea of
fermions of mass $m$ (the bare or band mass), and $\chi_0^\perp$ is the
transverse current-current response, of the same Fermi sea, including the 
constant ``diamagnetic current'' term. In the limit
where first the frequency $\omega$ and then the wavevector $q$ tend to
zero, we have 
\begin{eqnarray}
\chi_0&=&m/2\pi,\\
\chi_0^\perp&\sim& -q^2/12\pi m,
\end{eqnarray}
and hence
\begin{equation}
\frac{\partial n}{\partial \mu}\equiv\lim_{q\rightarrow 0}\chi^{\rm
irr}_{\rho\rho}(q,0)=\frac{m/2\pi}{1+\tilde{\phi}^2/6}.
\end{equation}
(For a long-range potential, i.e.\ one that is divergent as
$q\rightarrow0$, this is the appropriate definition of the
compressibility $\partial n/\partial \mu$. For a short range interaction,
one would use
$\chi_{\rho\rho}$ in place of $\chi_{\rho\rho}^{\rm irr}$.) Thus the
theory predicts that the system is {\em compressible}. Note however that
the approach describes the properties that the system has if it is in the
phase described. For a highly-correlated system such as particles in the
lowest Landau level, it is difficult to find any approach that can
accurately
predict, for a given Hamiltonian, in which phase the system will be. For
example, an alternative phase that is possible at the same filling factors 
as the Fermi liquid (FL) is the Pfaffian state \cite{mr}, which is
believed to be incompressible\cite{rr96}. Nevertheless, the question of
the properties of the Fermi liquid state---which has a Fermi surface in
the excitation spectrum for the fermions---is well-defined.

For the conductivity, the general statement \cite{il} is that the
resistivity tensors add:
\begin{equation}
\rho=\rho_{\rm CS}+\rho_\psi,
\label{iofflark}
\end{equation}
where $\rho_{{\rm
CS}\mu\nu}=2\pi\tilde{\phi}\varepsilon_{\mu\nu}$, coincides with the
Hall resistivity at $\nu=1/\tilde{\phi}$ and $\rho_{\psi\mu\nu}$ is the 
resistivity tensor of the
fermions, the inverse of the conductivity tensor which is related to the
current-current response function that is irreducible with
respect to both the interaction and the gauge field. In the RPA, using
the Drude approximation to include impurities, one has \cite{hlr} at
$q\rightarrow0$,
then $\omega\rightarrow0$, $\rho_{\psi\mu\nu}=
\delta_{\mu\nu}/\sigma_{\psi xx}$ where $\sigma_{\psi xx}$ is the usual
Drude result for the Fermi sea in zero magnetic field with impurity
scattering. There is also an unusual scattering mechanism \cite{hlr,kz} 
in which the fermions scatter off the static vector potential $\delta{\bf
a}$ induced in the Chern-Simons gauge field by a variation in the density
of particles produced by the impurity potential, since
$\nabla\wedge\delta{\bf a}=-2\pi\tilde{\phi}\delta\rho$. 

The effects of interactions and gauge field fluctuations beyond RPA would
be expected to have a variety of effects. By analogy with the Landau-Silin
treatment of fermions with a long range interaction, one would expect that
when both the long-range part of the interaction (if any) and of the
Chern-Simons gauge field are extracted, by considering responses
irreducible with respect to both the interaction and the gauge field as
above, the remaining effects
can be handled to all orders by renormalizing parameters, and the leading 
long-wavelength effects expressed in terms of Landau interaction
parameters $F_\ell$, and an effective mass $m^\ast$. Since the system is
translationally and Galilean invariant (in the absence of impurities), the
latter mass must satisfy the usual relation 
\cite{hlr,sh}
\begin{equation}
m^\ast/m=1+F_1
\end{equation}
(details of our two-dimensional normalization of the Landau parameters
such as $F_1$ are 
given later). In addition, in the limit where the cyclotron energy
$\omega_c=1/m$ is large compared with the typical interaction strength
between particles, $V(\bar{\rho}^{-1/2})$, (e.g.\ as $m\rightarrow0$), the
dynamics should be governed
entirely by the interactions, and so $1/m^\ast$ should scale with the
interaction strength, and be of order the typical interaction strength up
to numerical factors. 

This expectation that the theory would be a renormalized Fermi liquid,
coupled to the long-range interaction and the gauge field, turned out
to be too naive, however. The fluctuations of the gauge field have
singular effects that appear to cause a partial breakdown of the Fermi
liquid picture \cite{hlr}. The effects of such fluctuations were evaluated
in leading order in the RPA gauge field propagator in HLR (the small
parameter being $\tilde{\phi}$, with the background magnetic field being
adjusted such that the net field seen by the fermions on average was zero,
for any value of $\tilde{\phi}$, i.e.\ the filling factor was always
$1/\tilde{\phi}$; recall that for generic values of $\tilde{\phi}$ the
particles are anyons). The main effects were, first, that the propagator
itself shows the appearance of a mode at the cyclotron frequency
$1/m$, which carries all of the f-sum rule
spectral
weight to order $q^2$. Thus this mode is the physical cyclotron mode. 
The virtual excitation of this mode, which is the longitudinal part of
the gauge field, led, in first order, to a contribution
to the fermion self energy that was logarithmically infrared divergent.
The effect could plausibly be exponentiated to give for the quasiparticle 
residue $Z_F$ of a fermion at the Fermi wavevector $k_F$, 
\begin{equation}
Z_F\sim L^{-\tilde{\phi}/2},
\end{equation}
where $L$ is the system size (or, presumably, $|k-k_F|^{\tilde{\phi}/2}$
as $k$ approaches $k_F$ for infinite $L$). This would correspond to the 
Girvin-MacDonald (GM) power law \cite{gm}, generalized to the fermion
case; in particular, the exponent should be exact. This is supported by
further analysis of these fluctuations which, similarly to the boson case
\cite{kane}, lead to a factor $\prod_{i<j}|z_i-z_j|^{\tilde{\phi}}$, times 
a gaussian, in the ground state wavefunction of the fermions (the result
for the fermion case is widely known but does not appear to have been
explicitly published). This in turn leads to the GM power
$r^{-\tilde{\phi}/2}$ as a factor in the equal-time Green's function of
the fermion
\cite{mil}
\begin{equation}
\langle\psi({\bf r})\psi^\dagger({\bf 0})\rangle\sim
r^{-(3/2+\tilde{\phi}/2)} \sin(k_Fr-\pi/4) 
\end{equation}
and correspondingly to the above result for $Z_F$ (see also Ref.\
\cite{jr}). (The GM power law in the composite boson case has also been
recovered field theoretically in \cite{zhang}.) Related effects were also 
found in the work of Shankar and Murthy \cite{sm}, to which we shall turn 
shortly. In the work of HLR and others, it was assumed that the vanishing 
quasiparticle residue for the original CS fermions was of little 
significance, since as with many similar effects in field theory, in 
particular the nonsingular quasiparticle residue in an ordinary Fermi 
liquid, it cancels in physical response functions that measure 
quasiparticle properties. However, the recent results to be reviewed
below, and those of the present paper, suggest that things are not quite 
so simple, and rather than just ignoring these effects on the assumption 
that they cancel, the longitudinal mode should be integrated out 
``exactly'' to obtain an effective field theory, before proceeding to the 
effects of the other lower-energy fluctuations, such as the transverse 
fluctuations. 

The fluctuations in the transverse part of the gauge field have received
more attention (due to the CS term, there are also cross-terms that mix
the longitudinal and transverse fluctuations; however these are assumed
to have some intermediate significance). The first-order self energy
contains power-law infrared-divergent terms for the case of a short range
interaction, which are weakened by the presence of a long-range interaction
because the latter suppresses density fluctuations which correspond to
fluctuations of the transverse CS vector potential $\bf a$. For the $1/r$
Coulomb interaction, the effects become logarithmic, and for an
interaction which is longer range than $1/r$ they become finite. In the
Coulomb case, the structure of the effects is similar to those in an
electron gas coupled to the transverse part of the ordinary
electromagnetic field (since there is no CS term in this case, these
effects are not weakened by the Coulomb interaction, but are always
logarithmic---however, they are extremely weak in practise)
\cite{holstein,reizer}. In both of these systems, it can be argued by
treating the self energy self-consistently
\cite{holstein,hlr,polchinski} that the effects lead to an
effective mass diverging as $m^\ast\sim -\ln|k-k_F|$, a quasiparticle 
scattering rate
$\sim-|\varepsilon_k^\ast-\mu|/\ln|\varepsilon_k^\ast-\mu|$ 
(where $\varepsilon_k^\ast$ is the dispersion relation that corresponds to the
stated behavior of the effective mass near $k_F$), and a quasiparticle
residue $Z_F\sim-\ln|k-k_F|$ (the latter would be in
addition to the effect of the longitudinal fluctuations described above).
These results suggest that while the effective mass diverges at $k_F$, the
quasiparticles remain just marginally well-defined due to the
reciprocal logarithm in the decay rate, and thus the system is a
``marginal Fermi liquid''. For longer-range interaction, there is no such
breakdown of Landau Fermi liquid theory, and for the extreme case of
$V(r)\sim \ln r$, the scattering rate recovers its usual form $\sim
(\varepsilon_k-\mu)^2$ (all these results are for zero temperature). 

There are many other studies of this
\cite{naywil,khveshchenko,ganwan,ioffe,palee,sternhalp,houghton}, 
often with conflicting results.
We believe that the correct results are those that agree with the above
scenario of HLR for the behavior of the effective mass, etc. 

If we are not too concerned about the latter effects of transverse gauge
field fluctuations, for example if we consider an interaction longer-range
than Coulomb, or in the Coulomb case neglecting the logarithmic effects in
view of how slowly they diverge at $k_F$, then we are led to a physical
picture of what to expect {}from the system to all orders in the
fluctuations. It is essentially the Landau theory with due regard to the
long-range effects, as described above, and thus retains the CS structure
present in the RPA. For the density-density response,
the responses $\chi_0$ and $\chi_0^\perp$ that appeared in the RPA will
therefore be replaced by renormalized versions, and according to this
scenario, we then expect that, in the limit that gives (for example) the  
compressibility, $\chi_0$ and $\chi_0^\perp$ that appeared in the RPA will 
be replaced by renormalized versions, $m^\ast/[2\pi(1+F_0)]$ and $q^2
\chi_{\rm d}^\ast$ respectively, where $\chi_{\rm d}^\ast$ is a
renormalized long-wavelength Landau diamagnetic susceptibility, which is a
non-Fermi-liquid quantity  as it involves derivatives at the Fermi
surface. Explicitly, we expect
\be
\frac{\partial n}{\partial \mu}=\frac{m^\ast}{2\pi(1+F_0)-(2\pi\pt)^2
\chi_{\rm d}^\ast m^\ast}.
\label{cscomp}
\ee
(We expect that $F_0$ diverges the same way as
$m^\ast$ so that
the renormalized version of $\chi_0$ remains finite
\cite{palee,sternhalp}.) 
Thus the system remains compressible in this scenario. 

\subsection{Physical pictures}
\label{physical}

In this subsection, we review aspects discussed in Ref.\ \cite{read94},
which was in part an elucidation of Ref.\ \cite{read89} (see also
\cite{read95}). The approach is based on the wavefunctions of the
particles, which are assumed {}from the beginning to be in the lowest Landau
level (LLL). To lower the repulsive interaction energy, each particle
would like to bind to $\tilde{\phi}$ vortices, which at
$\nu=1/\tilde{\phi}$ leaves no vortices left. (Note that in the LLL, the
number of zeroes in the wavefunction of each particle is equal to the
number of flux quanta threading the relevant area, and that a vortex
means a simultaneous zero in the wavefunction of every particle other than 
the one under consideration.) For the same choices of statistics of the
particles and of filling factor as before, the bound states behave as
fermions in zero net magnetic field (this statement again involves the
mean field assumption that the average density of particles is
uniform, as we will see). 

To make the idea concrete, we may consider trial
wavefunctions in which the fermionic bound states occupy a Slater
determinant of plane waves, or spherical harmonics on the sphere
\cite{rr94} (these resemble Jain's trial wavefunctions \cite{jain}, except
that the fermions are in zero effective magnetic field)
\begin{equation}
\tilde{\Psi}(z_1,\ldots,z_N)={\cal P}_{\rm LLL}\det M_{ij}\,\prod_{i<j}
(z_i-z_j)^{\tilde{\phi}}.
\end{equation}  
Here we write the wavefunction on the sphere \cite{hald83}, with
$z_i=2Rv_i/u_i$ the complex coordinate of particle $i$ in stereographic
projection to the plane. Only the polynomial part of the wavefunction is
shown, as indicated by the tilde on $\Psi$. The full wavefunction is recovered 
by multiplication by $\prod_i(1+|z_i|^2/2R^2)^{-(N_\phi+2)/2}$, and this must 
be done before integration of the $z_i$ coordinates over the complex plane, 
to give the correct integration measure, in particular when applying the 
LLL projection operator ${\cal P}_{\rm LLL}$. In the limit where the 
radius $R$ and the number $N_\phi$ of flux quanta through the surface of 
the sphere go to infinity with the field strength fixed, the 
non-polynomial factor approaches the usual 
$e^{-\frac{1}{4}\sum_i|z_i|^2}$. $M_{ij}$ are the spherical harmonics of 
angular momentum $L_i$, $M_i$ for the $j$th particle, or can be replaced
by plane waves $e^{i{\bf k}_i\cdot{\bf r}_j}$ in the plane. The $L_i$,
$M_i$ 
(or ${\bf k}_i$) can be chosen to fill the Fermi sea to obtain a trial 
ground state \cite{rr94}. Different choices of the sets of $L_i$, $M_i$ do 
{\em not} give orthogonal states in general, except when the total angular 
momenta differ. Note that apart {}from the projection to the LLL, the
wavefunctions have the form that would be expected {}from the CS approach,
on including the fluctuations at the RPA level that produce the amplitude
of the Laughlin-Jastrow (LJ) factor in the wavefunction, as noted above. 

The fermionic bound states or ``quasiparticles'' described here are
created by operators of the form $\psi_e^\dagger U^{\tilde{\phi}}$, where
$\psi^\dagger_e$ creates a particle in the LLL, and $U(z)=\prod_i(z_i-z)$
is Laughlin's quasihole operator \cite{laugh}, which creates a vortex
\cite{read89}. As for the wavefunctions, this differs {}from the CS fermion
operator $\psi^\dagger$ by including the amplitude of the quasihole
operator, and not just the phase (like the wavefunctions, it should also
include a non-polynomial factor in $z$, which we have suppressed here). 
Consequently, like the corresponding boson operator \cite{read89}, its 
equal-time Green's function is not expected to include the GM power-law 
factor $r^{-\tilde{\phi}/2}$; this has been confirmed by calculation 
\cite{mil}. Since at $\nu=1/\tilde{\phi}$ the $\tilde{\phi}$ vortices
induce a hole in the density of the other particles that contains a
deficiency in the particle number of exactly unity, there has always been 
a temptation to say that the bound states formed this way are neutral
objects. This should be contrasted with the CS fermions and bosons, which
carry particle number unity. 

The plane-wave factors, in the flat space limit, can be rewritten using
\cite{gj,gmp} (see also Appendix \ref{noncom})
\be
{\cal P}_{\rm LLL}e^{i\bk\cdot\br_i}{\cal P}_{\rm LLL}=e^{i\bk\cdot{\bf
R}_i}e^{-\frac{1}{4}|k|^2},
\ee 
where ${\bf R}_i$ is the guiding-center coordinate of particle $i$, which
has no matrix elements between states in different Landau levels. The
operator ${\bf K}_i=-\wedge{\bf R}_i$ is the pseudomomentum that generates
magnetic translations of particle $i$. Thus, the plane-wave factors in the
Slater determinant can be replaced by $e^{i\bk\cdot{\bf R}_i}$ and each
such
factor displaces the $i$th particle by $\wedge\bk$ (in units where the
magnetic length is one) {}from its vortices. This picture of particles
bound to vortices but displaced by $\wedge\bk$ {}from their center has
several consequences \cite{read94}.

The first consequence is that, if we consider the interaction of the
particle with the
vortices (or correlation hole) to which it is bound (neglecting the
exchange effects due to the latter being constructed {}from other particles,
indistinguishable {}from the first), then for $\bk=0$, the particle is
precisely on the vortices as in the Laughlin states, and for $\bk\neq0$ it
is displaced by $\wedge\bk$. Consequently, the energy should increase, and
the interaction between the particle and its vortices becomes an effective
kinetic (i.e.\ $\bk$-dependent) energy for the fermion, which is the
origin of the effective mass at the Fermi wavevector, and scales inversely
with $V$. A formula for this energy can be found for the analogous
boson case in Ref.\ \cite{read89}. Notice that the displacements in the
Fermi sea ground state are bounded above by $k_F=\sqrt{2/\tilde{\phi}}$,
which is much less than the typical distance between neighboring particles 
which is of order $\sim\sqrt{\pt}$. Thus for $\pt>$ order 1, which is the 
case of interest when the particles are bosons or fermions, not anyons, 
the displacements do not unduly perturb the bound states.       

Second, if we accept that the fermions are neutral, then their leading
coupling to the electric potential is through a dipole moment
$\wedge\bk$. It is important to realise that the wavevectors of the
fermions contribute to the total momentum of the system, which is a
conserved quantity. One might imagine that the dipole moment could be
renormalized by effects not yet included, or that the vortices might not
all be at the same point as we have implicitly assumed. Indeed, when the
underlying particles are fermions, the wavefunction must have one vortex
exactly on every particle, because of antisymmetry. This will not affect
the dipole moment, because the plane-wave factors must produce the
displacement shown, and when the particles are fermions, this is
accomplished by displacing the other vortices further to compensate for
the one that is not displaced at all. Also, if the vortices are viewed as 
point objects, then their relative displacements can only produce
multipole moments of even order, and not a contribution to the dipole
moment, which is determined by the displacement of the particle {}from the
center of mass of the vortices. Thus the dipole moment is not
renormalized. A more rigorous version of this argument
will be given later in this paper.

Third, when the $\pt$ vortices are dragged around adiabatically, they pick
up a Berry phase factor \cite{asw} which can be interpreted as a vector and
scalar potential governed by the particle number and number drift-current
densities, $\rho$ and $\bf j$ \cite{read89,read94,read95}. This means that
the fermionic bound states experience, in addition to the electromagnetic
$\bf A$ and $A_0$, also $\ba$, $a_0$ given by 
\bea
\nabla\wedge\ba&=&-2\pi\pt\rho,\\
-\dot{\ba}-\nabla a_0&=&2\pi\pt\wedge{\bf j}.
\eea 
These have the form of the equations in the CS fermion approach, but it is
important to emphasise that they have been obtained
\cite{read89,read94,read95} without the use
of $\delta$-function fluxes attached to the particles, and that they still
involve the physical density and current, which cannot be identified with
the density and current of the fermions because the latter are (or 
may be) neutral.

In Ref.\ \cite{read94}, these were used as an alternative approach that
was stated to be equivalent to the CS approach, and the neutrality of the
quasiparticles was not invoked. It was felt that, although the fermions
and bosons appear neutral, the situation might be like that in the usual
electron gas problem with a Coulomb interaction, where at low energies the
quasiparticles are neutral in their couplings to external
longitudinal electric fields because of screening, but in the Fermi liquid
viewpoint, one nonetheless views the fermions as having charge unity, and
the low-energy behavior of the Fermi liquid itself produces the screening
effects, in the limit $\omega/q\rightarrow0$ in the response functions.
In the opposite limit $\omega/q\rightarrow\infty$, the charge of the 
quasiparticles does show up, in the conductivity (and also in the 
transverse response in both regimes). However, recent work to be discussed
in the next subsection, and the work in the present paper, suggests that 
in the quantum Hall effect context, we can in fact obtain a consistent 
picture in which the quasiparticles have only dipolar couplings to external 
fields. The obvious question is then whether the Fermi liquid is still 
compressible. We will answer this question in the affirmative.  

\subsection{Recent approaches to the LLL}
\label{recent}

Several recent works have taken up the outstanding issues discussed in the
previous subsections. They are concerned with obtaining results for
the Fermi-liquid state including the effects of all the particles being in
the lowest Landau level, or as would seem to be at least roughly
equivalent, including the effects of the amplitude of the correlation
factors produced by the zero-point fluctuations of the cyclotron-frequency
longitudinal modes of the CS gauge field. The aim of such work is, of
course, to test the validity of the results of HLR. Different approaches
have been used. Shankar and Murthy (SM) \cite{sm} base their work on the 
U(1) CS fermion field theory approach, however they work in a Hamiltonian
formalism, and aim to eliminate the cyclotron variables by canonical
transformation, rather than by resummation of perturbation theory. The
cyclotron modes are represented as oscillators whose zero-point motion
produces the amplitude of the LJ factor in the
ground-state wavefunction. However, when fermions are excited to different
$\bk$ states, the oscillators must adjust to a displaced ground state, and
this seems to reproduce many of the effects of the correlation hole 
discussed in the preceding Subsection, as well as other effects connected 
with the cyclotron mode and the projection to the lowest Landau level.
D.-H.~Lee \cite{dhlee} uses duality methods,
which are good for representing vortices. In his approach, the
particles are fermions at $\nu=1/2$, but, in view of the single vortex
exactly on each particle because of Fermi statistics (for LLL
wavefunctions), they can be represented as bosons at $\nu=1$. In these two
works, only the leading long-wavelength effects can be treated. Pasquier 
and Haldane (PH) \cite{ph} use a method that is valid only for $\pt=1$
(that is, the particles are bosons at $\nu=1$), and represents the
LLL problem exactly, through equations valid for all wavelengths. A
version of their method will be described in the next section and used
extensively in this paper. 
 
All these groups arrive at the following points in common. The LLL physics
is described by Fermi fields $c$, $c^\dagger$ with canonical anticommutation 
relations, and the physical states must obey the operator constraints for 
each wavevector $\bq$,
\bea
\lefteqn{\int \frac{d^2k}{(2\pi)^2}
c^\dagger_{\bk-\frac{1}{2}\bq}c_{\bk+\frac{1}{2}\bq}
(1-\frac{1}{2}i\bk\wedge\bq)}\non\\
&&\mbox{}+O(q^2)-\bar{\rho}(2\pi)^2\delta(\bq)
=0.
\label{smconstr}
\eea
In SM and Lee, the $O(q^2)$ terms are unknown, and in SM the constraints
are further restricted to apply only for $q$ less than a cutoff $Q$ whch
is chosen to equal $k_F$. In PH, the terms higher-order in $q$ are known. 
The physical particle number-density operator reduces to the form
\be
\rho(\bq)=
\bar{\rho}(2\pi)^2\delta(\bq)+\int \frac{d^2k}{(2\pi)^2} i
\bk\wedge \bq\,
c^\dagger_{\bk-\frac{1}{2}\bq}c_{\bk+\frac{1}{2}\bq},
\label{dipdens}
\ee
again to leading order in $q$, on using the constraints. Note that this
is the Fourier transform of a dipolar or polarization expression
for the density, $\rho=\bar{\rho}-\nabla\cdot {\bf P}$, where the
polarization ${\bf P}$ is that due to a dipole moment of $\wedge\bk$ on a
fermion of wavevector $\bk$ (this semiclassical way of describing it will
be quite useful; compare the discussion of fermions with a fairly
well-defined wavevector and position in Fermi-liquid theory, which can be 
better described formally by the Wigner distribution function).
Lee differs {}from the other authors and {}from Ref.\ \cite{read94} in finding
an extra factor of $1/2$ in the right-hand side of Eq.\ (\ref{dipdens});
the origin of this $1/2$ is not clear to us. 

A result for the effective mass was obtained as follows. Beginning {}from
the interaction Hamiltonian that is all that is left when the kinetic
energy of the particles has been quenched,
\be
H_{\rm int}=\frac{1}{2} \int \frac{d^2q}{(2\pi)^2} V(q)
:\rho(\bq)\rho(-\bq):
\ee
where colons $:\ldots:$ represent normal ordering, the normal ordering is
then dropped as it produces only a constant proportional to the number of
particles. The density is then replaced by the form in Eq.\
(\ref{dipdens}). When this is written in first quantization it becomes
\be
H_{\rm int}=\frac{1}{2}\sum_{ij} \int \frac{d^2q}{(2\pi)^2} V(q)
\bq\wedge\bk_i \bq\wedge\bk_j.
\ee
On taking the $i=j$ term of this expression, they obtain an effective
kinetic energy due to interactions,
\be
\sum_i \bk_i^2/(2m^\ast)
\ee
where the effective mass is given by
\be
1/m^\ast=\frac{1}{2}\int \frac{d^2q}{(2\pi)^2} V(q)q^2,
\ee
which has the form of the dipole-moment-squared term in
the self-interaction energy of a dipole, and if the $q$ integral is cutoff
as in SM, the density profile is smeared as it would be in the correlation
hole. It is therefore similar to the proposal of Ref.\
\cite{read89,read94}.   

For the density-density response function, these authors find, using
the dipolar form of the density,
\be
\chi_{\rho\rho}(\bq,0)=\langle\rho(\bq)\rho(-\bq)\rangle=q^2\langle 
PP\rangle=q^2\bar{\rho}m^\ast+ O(q^4).
\ee
In the last step, the transverse momentum-momentum response
function of the Fermi gas with effective mass $m^\ast$ was used. In these
calculations, the constraints (\ref{smconstr}) were either ignored
\cite{ph}, or
were handled by introducing functional-integral representations of
$\delta$-functions of the constraints, which were then treated in the RPA 
\cite{sm}; the results take the form stated in either case. 

If this last result is taken seriously, it implies that the system is
incompressible. However, SM state some reservations about the calculation,
because of the way the constraint was handled. They suggest that the
symmetry of the Hamiltonian under translations of the wavevectors of all
the particles could lead to cancellations and to factors of $1/q^2$ that
could restore a finite compressibility to the system. This proposal is
very close to the results that will be obtained in the present paper by a
systematic treatment of the constraints. While this paper was being
completed, a short comment \cite{halpsterncomm} and a revised version of
Ref.\ \cite{dhlee} appeared which use the same symmetry just mentioned and
obtain results very
similar to some of ours below, including the fact that the system is
compressible. We will comment further on the relation of the symmetries
being used in Sec.\ \ref{allorders}.

\subsection{Overview of the results of the present paper}
\label{overview}

Here we describe results of the present paper. First we give a simple
discussion of our central result, for the density-density response
function. With the benefit of hindsight, using arguments that are justifed
by the more detailed and formal calculations below, the results can in
fact be obtained {}from the results of Sec.\ \ref{physical}. 
Then we describe the results of this paper.

The dipolar form of the density in Sec.\ \ref{physical} can be expressed
as 
\be
\rho(\br)=\bar{\rho}-\nabla\wedge{\bf g},
\ee
where ${\bf g}(\br)$ is the momentum density of the fermions, since ${\bf
P}=\wedge{\bf g}$. On the other hand, we also have
\be
\rho(\br)=\bar{\rho}-\bar{\rho}\nabla\wedge(\ba+\bA).
\ee
This suggests that we write 
\be
\ba+\bA={\bf g}/\bar{\rho}
\label{aequalsg}
\ee
in general, even though the above argument only implies this for the
transverse part of $\ba$. This equation suggests there is a
gauge-invariant current ${\bf j}^R$, which is not the physical current, 
such that (for excitations near the Fermi surface), 
\bea
{\bf j}^R&=&\left\{-i\frac{1}{2}[c^\dagger\nabla c-(\nabla
c^\dagger)c]-(\ba+\bA)c^\dagger c\right\}/m^\ast\non\\
&=&[{\bf g}(\br)-(\ba+\bA)\rho^R]/m^\ast,
\eea
which is required to vanish, ${\bf j}^R=0$. Assuming that the ``density''
$\rho^R=c^\dagger c$ is just $\bar{\rho}$, this is equivalent to Eq.\
(\ref{aequalsg}). Indeed, vanishing current would be consistent with such
a constraint, $\rho^R=\bar{\rho}$, if they together obey a continuity
equation,
\be
\partial \rho^R/\partial t +\nabla\cdot{\bf j}^R=0.
\ee
This involves the longitudinal part of the current, so we have an argument
for both parts of Eq.\ (\ref{aequalsg}). The condition $\rho^R=\bar{\rho}$
should of course be viewed as the long-wavelength version of the
constraint found by SM, Lee, and PH. 

The gauge-invariant form of the ``current'' ${\bf j}^R$ encourages us
to consider an effective Hamiltonian
\be
H_{\rm eff}=\frac{1}{2m^\ast}\int d^2r\,|(-i\nabla-\ba-\bA)c|^2+\ldots,
\label{ourHeff}
\ee
which, apart {}from higher covariant derivatives of $c$, $c^\dagger$,
contains no other terms in $\ba$, not even a CS term. Thus $\ba$ is
a strongly-coupled gauge field and varying $H_{\rm eff}$ with respect to
$\ba$ yields $\bj^R=0$. Then, neglecting other terms in $H_{\rm eff}$, we
can use the RPA, or the following
self-consistent field argument, to find the density-density response
function. {}From the form of the density, an external scalar potential
couples to $\nabla\wedge\bg$. The irreducible density response
contains two
parts. The first part is {}from the transverse momentum-momentum response
function of the gas with mass $m^\ast$; it is the part found by SM, Lee,
and PH \cite{sm,dhlee,ph} (Lee has since revised this result
\cite{dhlee}). The second is the response of the
same gas to the induced vector potential $\ba$. (In both responses, the 
constant ``diamagnetic current'' term is absent.) Thus
\be
\chi_{\rho\rho}^{\rm irr}=(\bar{\rho}+m^\ast\chi_0^\perp)(q^2
m^\ast+iq\delta a_\perp)
\ee
where in the last factor the two terms arise {}from the two parts just
described, and $\delta a_\perp$ is the response in the transverse vector
potential to the perturbation, and is therefore given by
\be
iq\delta a_\perp=\chi_{\rho\rho}^{\rm irr}/\bar{\rho}.
\ee
{}From these self-consistent equations we find
\be
\chi_{\rho\rho}^{\rm
irr}=-\bar{\rho}(\bar{\rho}+m^\ast\chi_0^\perp)q^2/\chi_0^\perp,
\ee
which is exactly the result we obtain in this paper.
This yields for the compressibility $dn/d\mu=-\bar{\rho}^2/\chi_{\rm
d}^\ast>0$, where $\chi_{\rm d}^\ast$ is the diamagnetic susceptibility 
for this fermion gas. This result differs {}from that in the scenario 
based on the U(1) CS approach, described at the end of Subsec.\
\ref{chern}. Several other observables
are similarly in close, but not always exact, agreement with the scenario
based on HLR, described above.

In this argument, we neglected the Landau parameters. These can be
included without significantly changing the results. However, the Landau
parameter $F_1$ should not be added, since it is already included in the
gauge field effects. {\em The strongly-coupled gauge field in the Fermi
liquid is equivalent to a Landau parameter $F_1=-1$}, provided $m^\ast>0$. 
Thus we are led to a scenario in which the Fermi-liquid-like state has
many FL properties in common with the theory of HLR, including 
a finite compressibility, yet differs in that there is no CS term for the
gauge field, while the physical density is dipolar or (using an equation
of motion) is $-\bar{\rho}\nabla\wedge\ba$.

In the rest of the paper, we follow a different argument {}from that just
presented. We give a detailed microscopic derivation, in which the
relationship $\rho(\br)=-\bar{\rho}\nabla\wedge\ba$ appears only at the
end; thus we do not rely on the Berry phase argument. The starting point
is an approach of Pasquier and Haldane, described in Sec.\
\ref{PHapproach} below. In this approach, which works for $\pt=1$ only,
that is bosons at $\nu=1$, each fermion is described by two coordinates, 
which we term ``left'' and ``right'', but the available states are those
of a particle in zero magnetic field, because the wavefunctions are
complex analytic in the left and antianalytic in the right coordinates.
The left coordinate is that of the underlying particle contained in the
fermion, while the right coordinate represents an attached vortex, as in
the pictures in Sec.\ \ref{physical}. The system must obey a constraint of
fixed density $\rho^R=\bar{\rho}$ in the right coordinates. Since the
separation of the left {}from the right coordinate is $\wedge\bk$ when the
fermion is in a plane wave state of wavevector $\bk$, the physical density
is dipolar. In order to maintain the constraint, the longitudinal part of
the current $\bj^R$ of the vortices (right coordinates) must vanish, as
argued above. In Sec.\ \ref{HFconserving}), we consider a conserving
approximation for observable response functions. We show that the
constraints are satisfied in this method. We calculate the density-density
response, its spectral density, the longitudinal conductivity, the
scattering of the fermions by a potential, and the current-density
operator. {}From the results we deduce that the system can be described in 
terms of the strongly-coupled gauge field mentioned above. The gauge
invariance is a manifestation of the constraint. The gauge fields obey the
CS {\em equations}, even though there is no CS {\em term} in the action.
In Sec.\ \ref{allorders}, we indicate the form we expect for the exact
results to all orders in the interactions, and give arguments why these
are correct. We conjecture that a certain sum rule for the spectral
density is exact. While at present this approach works for $\pt=1$, that
is for bosons at $\nu=1$, we expect that the conclusions are more general,
as the results and arguments of the previous subsections and the beginning
of this one are.

\section{Pasquier-Haldane approach\\ for $\pt=1$}
\label{PHapproach}

In this section we review (with a few variations of our own) the method of
PH \cite{phvar} which works only for $\pt=1$, though the filling factor
does not necessarily have to be one. A similar method works for fermions
with one vortex attached, mapping them to composite bosons. Since the
formalism has not appeared elsewhere in the form in which we will use it,
it will be presented in self-contained fashion.

We begin abstractly, labelling arbitrary single-particle states with
indices. Hopefully the later development in coordinate space, though less
general, will seem less abstract and give more physical insight, and show
clearly the connection with composites particles and the LLL. We take
fermion operators which are matrices with two indices, $c_{mn}$ and
$c_{nm}^\dagger$, with canonical anticommutation relations
\be
\{c_{mn},c_{n'm'}^\dagger\}=\delta_{mm'}\delta_{nn'}
\ee
(and others vanish)
where $m$, $m'$, $n$, $n'$ run {}from $1$ to $N$ (this case of square
matrices is convenient for the $\nu=1$ boson problem, while rectangular
matrices would be used for $\nu\neq 1$). 
The anticommutation relations are invariant under independent unitary
transformations on the left and right indices, under which
\bea
c&\mapsto&U_L c U_R,\non\\
c^\dagger&\mapsto&U_R^\dagger c^\dagger U_L^\dagger,
\eea 
where $U_L$, $U_R$ are unitary $N\times N$ matrices. These transformations
are generated by the operators
\bea
\rho_{nn'}^R&=&\sum_m c_{nm}^\dagger c_{mn'},\\
\rho_{mm'}^L&=&\sum_n c_{nm'}^\dagger c_{mn}.
\eea
The right generators $\rho^R$ generate the group U($N$)$_R$ of unitary
matrices. These are used to specify a set of $N^2$ constraints on the
system,
\be
(\rho_{nn'}^R-\delta_{nn'})|\Psi_{\rm phys}\rangle=0
\ee
which defines a subspace of states that will be identified with
the physical Hilbert space.
By taking the trace, we see that these imply that the U(1) generator
or fermion number operator (which is common to U($N$)$_R$ and U($N$)$_L$) 
\be
\hat{N}=\sum_{mn}c_{nm}^\dagger c_{mn},
\ee  
must have eigenvalue equal to $N$.
Thus in the allowed subspace, $N$ is both the range of the indices, and
the number of fermions. The remaining right generators generate
SU($N$)$_R$, and physical states must be singlets under the action of this
group. The other group, SU($N$)$_L$, is not used for constraints, and will
be broken by the Hamiltonian to a subgroup that represents translations
and/or rotations on the two-dimensional manifold (say, the sphere,
torus, or infinite plane) on which the physical
particles move. At the same time, the generators $\rho_{mm'}^L$ will
represent the physical density on this manifold.

The physical states that satisfy the constraints can be written as linear
combinations of 
\be
|\Psi_{\rm phys}^{m_1\ldots m_N}\rangle =
\sum_{n_1,\ldots,n_N}\varepsilon^{n_1\ldots n_N}
c_{n_1m_1}^\dagger c_{n_2m_2}^\dagger\cdots  c_{n_N m_N}^\dagger|0\rangle,
\ee
where $|0\rangle$ is the vacuum containing no fermions. These states
contain $N$ fermions and are clearly singlets under SU($N$)$_R$ since they
are antisymmetric in the $n$ (right) indices. On the other hand, 
the anticommutation of the $c^\dagger$'s implies that they are symmetric
in the remaining $m$ (left) indices. Thus these states can be viewed as
basis states for a system of $N$ bosons, each of which can be in any one
of $N$ single-particle states. Such a boson system could be described by
basis states
\be
a_{m_1}^\dagger\cdots a_{m_N}^\dagger|0\rangle,
\ee
where $[a_m,a_{m'}^\dagger]=\delta_{mm'}$ and others vanish. Each such
state is obtained in this way, which proves that the fermion system of
$c's$ with the constraints is equivalent to the unconstrained boson
system. If we define a filling factor as the 
particle number divided by the number of available orbitals, as
$N\rightarrow\infty$, then in our
case we clearly have bosons at filling factor $\nu=1$.

We note that in the larger Hilbert space without the constraints,
which is just the Fock space of the $c$'s, each fermion can be in any of
$N^2$ states, so there are 
\be
{N^2\choose N}
\ee
linearly-independent states for $N$ fermions. The states satisfying the
constraints form the Fock space of the bosons $a$, which contains only 
\be
{2N-1\choose N}
\ee
linearly-independent states.  

The left indices $m$ can run over any range, and this can be
used to represent any filling factor $\nu$. The constrained system can
also be set up using canonical {\em commutation} relations for the $c$'s,
and a similar argument then shows that the physical states
represent {\em fermions}, (e.g.\ electrons) at $\nu\leq 1$. 

So far we have only a way of representing bosons by fermions (or vice
versa), and the technique is reminiscent of the methods used for quantum
spin systems, in the case of a single quantum spin (see e.g.\ \cite{rs}). 
If we now view $m$, $n$ as indices for lowest Landau level states, 
say on a sphere where there are $N_\phi+1$ such states for $N_\phi$ flux
quanta through the sphere \cite{hald83}, then for the case where both
indices range
{}from $1$ to $N$, we have $N=N_\phi+1$, and the filling factor $\nu$ agrees
with that defined as $N/N_\phi$ as $N\rightarrow\infty$. We can introduce
coordinate space wavefunctions for the left index $m$, which are just
those of the physical bosons. We do the same for the right indices $n$,
except that they are complex conjugated so that the field strength (or
the charge) is effectively reversed. Using orthonormal single-particle
LLL basis states $u_m(z)$, we write in analogy with the usual field
operators
\bea
c(z,\wb)&=&\sum_{mn}u_m(z)\overline{u_n(w)}c_{mn},\non\\
c^\dagger(w,\zb)&=&\sum_{mn}u_n(w)\overline{u_m(z)}c_{nm}^\dagger,
\eea
which are adjoints of each other. Note that we use $z$'s for ``left''
indices, corresponding to $m$'s (which however appear on the right in
$c^\dagger$) and $w$'s for ``right'' indices, corresponding to $n$'s. 
The appearance of two coordinates on $c$ and $c^\dagger$ means that they
behave like operators on the LLL single-particle Hilbert space, just like
the matrix structure they had in index notation. A formalism for 
handling such operators as integral kernels is given in Appendix
\ref{noncom}.
For the sphere, we can write $\tilde{u}_m(z)\propto z^m$, for $m=0$,
\ldots, $N_\phi=N-1$, and the factor $(1+|z|^2/4R^2)^{-(N_\phi+2)/2}$ must 
be attached before integration. Following this convention we will write
only the polynomial part in the following wavefunctions. 

In the $z$, $w$ variables, the densities become
\bea
\rho^R(w,\wb')&=&\int d^2z\, c^\dagger(w,\zb)c(z,\wb'),\\
\rho^L(z,\zb')&=&\int d^2w\, c^\dagger(w,\zb')c(z,\wb).
\eea
Matrix multiplication has been replaced by integration, so that all
operators in the single-particle Hilbert space of LLL functions of
$z$ and $\wb$ become integral kernels (see Appendix \ref{noncom}). One can
see that
$\rho^L(z,\zb)$
is the LLL-projected density operator denoted $\bar{\rho}$ by Girvin,
MacDonald and Platzman (GMP)
\cite{gmp}, and $\rho^R$ is analogous.

Passing to the thermodynamic limit at fixed field strength and
density $=\bar{\rho}$, the radius of the sphere goes to infinity, the
system becomes flat locally, and we may use Fourier transforms. The
version of the Fourier transform required is defined in Appendix
\ref{noncom}. To
avoid discussion of global issues, which would distinguish this
thermodynamic limit {}from that of a torus, we will view the use of Fourier 
transforms as a technique for handling local calculations, in which we 
could include damping factors which tend to unity at the end. 
Alternatively, every calculation could, with only a little extra difficulty, 
be done in coordinate space. A third alternative would be to use the analog of
the Fourier transform, involving spherical harmonics, on the finite size
sphere. This is more tedious. Introducing the Fourier transform in the
plane, then, we notice that the pair of coordinates $z$, $\wb$ for each 
particle or field operator $c$ is replaced by a single ordinary 
two-dimensional wavevector $\bk$. This makes sense because, by choosing equal 
and opposite field strengths for the basis functions in these coordinates, the
particles effectively see zero magnetic field, for our filling factor
$\nu=1/\pt=1$. Note that, because the functions are analytic in $z$,
$\wb$ (the LLL restriction), we do not have effectively four real
variables per particle, as we would if the basis states had not been 
restricted to the LLL. The transformation of the matrix $c(z,\wb)$ into a 
plane wave operator is similar to that for the density operator, say 
$\rho^L$, which can clearly be traded for its Fourier components (see 
e.g.\ GMP). 

In terms of $c_\bk$, $c_\bk^\dagger$, which are defined in Appendix
\ref{noncom}, 
and which satisfy
\be
\{c_\bk,c_{\bk'}^\dagger\}=(2\pi)^2 \delta(\bk-\bk'),
\label{cananti}
\ee
we have 
\bea
\rho^R(\bq)&=&\int\frac{d^2k}{(2\pi)^2}\,e^{-\frac{1}{2}i\bk\wedge\bq}\,
c_{\bk-\frac{1}{2}\bq}^\dagger
c_{\bk+\frac{1}{2}\bq}\label{rhoRfour},\\
\rho^L(\bq)&=&\int\frac{d^2k}{(2\pi)^2}\,e^{\frac{1}{2}i\bk\wedge\bq}\,
c_{\bk-\frac{1}{2}\bq}^\dagger c_{\bk+\frac{1}{2}\bq},
\label{rhoLfour}
\eea
and we can show that 
\bea
\left[\rho^R(\bq),\rho^L(\bq')\right]&=&0,\\
\left[\rho^R(\bq),\rho^R(\bq')\right]&=&-2i\sin\frac{1}{2}\bq\wedge\bq'
\,\rho^R(\bq+\bq'),\\
\left[\rho^L(\bq),\rho^L(\bq')\right]&=&2i\sin\frac{1}{2}\bq\wedge\bq'
\,\rho^L(\bq+\bq').
\label{Winfty}
\eea
The Lie algebra commutation relations defined by Eq.\ (\ref{Winfty}) 
appeared in GMP and in Ref. \cite{fairlie}, and the algebra
so-defined has become known as $W_\infty$ (the defining relations are
often given in a different basis of the Lie algebra, essentially the
expansion of our $\rho^L(z,\zb')$ in angular momentum eigenstates $z^m$,
$\zb'^{m}$). In the notation of GMP, our
$\rho^L(\bq)=e^{\frac{1}{4}|q|^2}\bar{\rho}(\bq)$.
(The following algebraic comments will not be used in the following.)
{}From our point of view, $W_\infty$ is just a certain limit of SU($N$) as
$N\rightarrow\infty$. It is also helpful to note that if the $2\sin
\frac{1}{2}\bq\wedge\bq'$ is replaced by $\bq\wedge\bq'$ in the
commutation relations (for example, because $\bq$, $\bq'$ or the magnetic
length are small), then the resulting algebra is that of ``area-preserving
diffeomorphisms'', or equivalently (for the corresponding Poisson bracket
relations) Fourier components of functions on classical phase space.
$W_\infty$ can then be viewed as a quantum deformation of the latter, thus
as ``diffeomorphisms of the quantum analogue of phase space'', a fairly
familiar view of the LLL. The connection of $W_\infty$ with the quantum
Hall effect has often been remarked \cite{winf}. Our interest here is in
the isomorphic algebra generated by the $\rho^R$'s, which are the constraints
of our problem.

The constraints become
\be
(\rho^R(\bq)-\bar{\rho}(2\pi)^2\delta(\bq))|\Psi_{\rm phys}\rangle=0. 
\ee
Thus states can be built up in the ``big'' Hilbert space as combinations
of
\be
\prod_{\{\bk_i\}}c_\bk^\dagger|0\rangle
\label{nkestates}
\ee
(where the product is indexed by $\bk$'s in a set of $N$ wavevectors
$\bk_i$), and then projected to satisfy the constraints. The effect of
projection can be more easily appreciated in terms of
wavefunctions in coordinate space, by
returning to the finite size system.

In coordinate space, the constraints require that the $\wb$ dependence 
of wavefunctions be that of a full LLL,
\be
\tilde{\Psi}_{\rm phys}(z_1,\wb_1,\ldots,z_N,\wb_N)=
   f(z_1,\ldots,z_N)\prod_{i<j}(\wb_i-\wb_j),
\ee
because the LJ factor in the $\wb$'s is the unique totally-antisymmetric
function annihilated by the $\rho^R$'s, since the full LLL has no density
fluctuations. Hence $f$ is a symmetric polynomial in the $z_i$'s, as
appropriate for
bosons. Projection of the wavefunction of any state in the ``big'' Hilbert
space to this physical subspace, where states can be characterized just by
$f$, is accomplished by multiplying by $\prod(w_i-w_j)$ and integrating
over the $w_i$'s with the appropriate measure, leaving a symmetric
function $f$ in the $z_i$'s (possibly zero). If as a family of examples we
take the states (\ref{nkestates}), or their analogues on the sphere,
in first quantization they become Slater determinants
$\det(Y_{L_iM_i}(z_j,\wb_j))$, where the $Y_{LM}(z,\wb)$ are spherical
harmonics projected to the LLL, which correspond to the plane waves
$\tau_\bk$ in the plane, defined in Appendix \ref{noncom}.
Then the projection gives
\bea
f&=&\int\prod_k d^2w_k \prod_{i,j}(w_i-w_j) \det
Y_{L_iM_i}(z_j,\wb_j)\non\\
&=&{\cal P}_{\rm LLL} \det Y_{L_iM_i}(\Omega_j)\prod(z_i-z_j),
\label{projtoRR}
\eea
that is, the projection to the LLL of ordinary spherical harmonics in a
Slater determinant times the LJ factor. These are just the
trial wavefunctions described in Sec. \ref{physical}. Thus the formalism
not only describes bosons at $\nu=1$, but the fermions are closely related
to those in the ``physical'' approach, where the amplitude of the
LJ factor is automatically included in the trial
wavefunctions. Contrast this with the CS approach, where the trial
wavefunctions satisfying the CS constraint of one flux attached to each
particle consist of the Slater determinant times only the phase of the LJ
factor, and no LLL projection. Note also that while the projection into a
strictly smaller subspace implies that states described by distinct sets
of $\bk_i$ before projection may not be orthogonal after projection, they
do not usually vanish, except in some exceptional cases noted in Ref.\
\cite{rr94}.

Since the right coordinates $\wb$ of the fermions become, in the trial
wavefunctions after projection, the locations of
the vortices, it seems natural to refer to them as such even before
projection. Thus we can say that {\em each fermion consists of a particle
(boson) at the left coordinate $z$, and a vortex at the right coordinate
$\wb$, and so as a whole is effectively neutral}. The constraints demand
that the density $\rho^R$ of vortex coordinates is fixed, as an operator
statement. 
This seems natural if the vortices are thought of as forming a
two-dimensional plasma (in view of the LJ factor and Laughlin's plasma
mapping \cite{laugh}), since the plasma is in a screening phase and
suppresses long-wavelength density fluctuations; indeed, in this case of
$\nu=1$, there are no fluctuations in the LLL density at all in the
Laughlin state (the full LLL or Vandemonde determinant). In retrospect,
this condition on the vortices seems to be the main effect that was left
out in Refs.\ \cite{read89,read94}.

Now we finally specify the Hamiltonian appropriate to bosons in the LLL at
$\nu=1$. In terms of the boson operators $a$ introduced earlier, we have,
assuming a potential interaction between the bosons, 
\be
H=\frac{1}{2}\sum_{m_1,\ldots,m_4}V_{m_1m_2;m_3m_4}a_{m_1}^\dagger
a_{m_2}^\dagger a_{m_4}a_{m_3},
\label{bosham}
\ee
where the matrix elements of the interaction in the LLL are \cite{fetter}
\bea
V_{m_1m_2;m_3m_4}&=&\int
d^2r_1\,d^2r_2\,\overline{u_{m_1}(z_1)}\overline{u_{m_2}(z_2)}\non\\
 &&\times  V(\br_1-\br_2)u_{m_3}(z_1)u_{m_4}(z_2).
\eea
The corresponding operator in the large Hilbert space,
where it commutes with the constraints $\rho^R$, and so projects to $H$ in
Eq.\ (\ref{bosham}), is 
\be
H=\frac{1}{2}\sum_{\stackrel{\scriptstyle
m_1,\ldots,m_4}{n_1,n_2}} V_{m_1m_2;m_3m_4}
c_{n_1m_1}^\dagger c_{n_2m_2}^\dagger c_{m_4n_2}c_{m_3n_1}.
\ee
Then using the definition of $c(z,\wb)$ we obtain
\be
H=\frac{1}{2} \int d^2r_1 d^2r_2 V(\br_1-\br_2)
:\rho^L(z_1,\zb_1)\rho^L(z_2,\zb_2):,
\ee
where the normal ordering is with respect to the vacuum of the $c$'s,
$|0\rangle$. Thus this is simply a potential interaction written in terms
of the LLL-projected density $\rho^L$. In Fourier space this becomes
\be
H=\frac{1}{2}\int \frac{d^2q}{(2\pi)^2}
\tilde{V}(\bq):\rho^L(\bq)\rho^L(-\bq):,
\label{haminrhoL}
\ee
where $\tilde{V}(\bq)=e^{-\frac{1}{2}|q|^2}V(\bq)$ absorbs a factor left
{}from the definition of the Fourier transform of $\rho^L$, and $V(\bq)$ is
the usual Fourier transform
\be
V(\bq)=\int d^2r\, e^{-i\bq\cdot\br}\,V(\br).
\ee
The interaction Hamiltonian breaks the symmetry group {}from SU($N$)$_L$ (in
the absence of interacton) to SU($2$) (for the sphere) or to magnetic
translations and rotations in the case of the plane. It still commutes
with the ``constraint operators''
\be
G(\bq)\equiv \rho^R(\bq)-\bar{\rho}(2\pi)^2\delta(\bq).
\ee

Our expression for the Hamiltonian differs somewhat {}from that in the paper
of PH. They work on the torus, which is a relatively unimportant
difference, and write the Hamiltonian using the
constraints to make the ansatz explained in Sec.\ \ref{recent}, which
results in a one-body term that gives the fermions an effective kinetic
energy coming {}from the interaction. In our approach we do not wish to
make such a substitution since the commutator of $H$ with the $G(\bq)$
would not vanish identically, but only on using the conditions $G(\bq)=0$.
The reason for our insistence on retaining $[H,G(\bq)]=0$ will be
discussed in the next section. Of course, if everything is done correctly,
the results should be the same, in the end, since the starting point is
the same. 

\section{Hartree-Fock and Conserving Approximations}
\label{HFconserving}

In this section, which is the central one of the paper, we develop an
approximate solution for our system that
descibes the FL state. We begin in Subsec.\ \ref{HF} with the
Hartree-Fock approximation, which yields a dispersion relation for the 
fermions. Then in Subsec.\ \ref{constraints} we explain how the
constraints can be included. We choose a gauge such that for nonzero
frequencies, they must be satisfied without any assistance {}from
integration over auxiliary fields that impose them explicitly. This is
achieved in Subsec.\ \ref{conserving} by use of conserving approximations,
a familiar method of many-body and quantum field theory. In the present
case, such an approximation consistent with HF is the generalized or
time-dependent HF approximation, which sums ring and ladder diagrams. We
show explicitly that the constraints are obeyed in our approximation. In
Subsec.\ \ref{asymptotics} we investigate the asymptotics of the ladder
series that appears in Subsec.\ \ref{conserving}, for use in the following
calculations. In Subsec.\ \ref{physresp} we apply the approach to the
physical response functions, beginning with the density-density response. 
We show that the system is compressible and that the longitudinal
conductivity relevant for the surface acoustic wave experiments, which is
a certain limit of this response, is given by exactly the same expression
as in HLR. We also exhibit a sum-rule-like relation for the high frequency
response, or for the first moment of the spectral density, which we will
later argue is exact. We consider the scattering of a fermion by a scalar
potential perturbation, and interpret the result in terms of a vector
potential related to the density by the CS relation discussed in Sec.\
\ref{review}. We calculate the longitudinal conductivity due to
impurity scattering. Finally, we consider the physical current density,
which we relate to the stress or momentum flux tensor of the fermions, and
so recover the other CS relation. 

\subsection{Hartree-Fock approximation}
\label{HF}

In this subsection, we use the HF approximation, which is quick and is 
the simplest one that gives an effective kinetic energy and is consistent
with a stable Fermi sea as the ground state. The treatment of the
constraints will be extensively discussed in the next subsection, and the
formalization of the exchange part of the self energy as the saddle point
approximation to a functional integral, valid in some sense in a large-$M$
limit (in a generalization of the model to $M$ component fermions), is
left to Appendix \ref{hubb}. 

The problem for $\pt=1$ using the PH approach is described by the
Hamiltonian (\ref{haminrhoL}) which can be written
\bea
H&=&\frac{1}{2}\int\frac{d^2k_1d^2k_2d^2q}{(2\pi)^6}\tilde{V}(q)
e^{\frac{1}{2}i\bk_1\wedge\bq-\frac{1}{2}i\bk_2\wedge\bq}
\non\\
&&\times
c_{\bk_1-\frac{1}{2}\bq}^\dagger c_{\bk_2+\frac{1}{2}\bq}^\dagger 
c_{\bk_2-\frac{1}{2}\bq}c_{\bk_1+\frac{1}{2}\bq},
\label{phHam}
\eea
subject to the constraints
$G(\bq)\equiv\rho^R(\bq)-\bar{\rho}(2\pi)^2\delta(\bq)=0$, that is
$\hat{N}=N$, and 
\be
\int\frac{d^2k}{(2\pi)^2}e^{-\frac{1}{2}i\bk\wedge\bq}
c_{\bk-\frac{1}{2}\bq}^\dagger c_{\bk+\frac{1}{2}\bq}=0
\ee
for $\bq\neq0$. Notice that when the phase factor containing
$\bk\wedge\bq$ is expanded in a Taylor series, to $O(q^2)$ it takes the
same form as the constraint found by SM and Lee \cite{sm,dhlee}, as
mentioned in Sec.\ \ref{recent}. 

The HF approximation for a translationally-invariant system
takes the energy eigenstates to be Slater determinants of plane waves,
that is plane-wave-occupation-number eigenstates in the second-quantized
formalism and the energy of such a state to be the expectation value of
$H$. As is well-known, for the excitation spectrum, this is equivalent to
replacing $H$ by an effective one-body Hamiltonian with an effective
energy $\varepsilon_\bk$ for each plane wave state $\bk$, where
$\varepsilon_\bk$ depends self-consistently on the occupation numbers
$n_\bk$. In the present case, we must also include the constraints by the 
use of Lagrange multipliers $\bar{\lambda}_\bq$ and minimise 
\be
H-\mu N-\int \frac{d^2q}{(2\pi)^2}\bar{\lambda}_\bq G(-\bq)
\ee
with respect to $\bar{\lambda}_\bq$ to find the ground state. When almost
all particles are in the Fermi sea, the $\bar{\lambda}_\bq$ are zero by
translational symmetry, except at $\bq={\bf 0}$ where $\bar{\lambda}_{\bf
0}$ absorbs the chemical potential $\mu$, consistent with the fact that
the constraints fix the particle number and hence we are actually in the
canonical, not grand canonical, ensemble. 
Consequently one has $\bar{\lambda}_\bq=(2\pi)^2\bar{\lambda}\delta(\bq)$, 
and $\bar{\lambda}+\mu$ is determined by the condition on the total
particle number. One arrives therefore at the total energy expectation
value,
\be
E=\frac{1}{2L^2}\sum_{\bk\bk'}f_{\bk\bk'}n_\bk n_{\bk'},
\ee
(in which we have used the conventional notation for a finite system in a
square box of side $L$, with discrete $\bk$ values, and $n_\bk$ are the
expectation values of the occupation numbers for the corresponding  
states), where 
\be 
f_{\bk\bk'}=\tilde{V}({\bf 0})-\tilde{V}(\bk-\bk').
\label{fkk'}
\ee
The function $f_{\bk\bk'}$ plays the role of the Landau interaction
function when $\bk$ and $\bk'$ are restricted to the Fermi surface.
The effective single-particle Hamiltonian $K=H-(\mu+\bar{\lambda})N$
is 
\be
K_{\rm eff}= \sum_\bk \xi_\bk c_\bk^\dagger c_\bk,
\label{Keff}
\ee
where $\xi_\bk=\varepsilon_\bk-\mu-\bar{\lambda}$ and
\be
\varepsilon_\bk=\tilde{V}({\bf 0}) \int \frac{d^2k'}{(2\pi)^2} n_{\bk'}^0
- \int \frac{d^2k'}{(2\pi)^2}\tilde{V}(\bk-\bk') n_{\bk'}^0,
\ee
in which the first term is the direct or Hartree term, equal to
$\tilde{V}({\bf 0})\bar{\rho}$,  and the second is
the exchange or Fock term, which is responsible for the $\bk$-dependence
of $\xi_\bk$. Also, in the ground state at zero temperature,
$n_\bk^0=\theta(k_F-k)$,
$k_F=\sqrt{2}$ in our units, and $\mu+\bar{\lambda}$ is chosen so that
$\xi_{k_F}=0$. Notice that the phase factors in the Hamiltonian $H$ have
turned out to be unity in the HF expressions, which are identical to those 
of the usual Fermi gas, except that the bare kinetic 
energy is zero, and that $\tilde{V}(\bq)$ replaces $V(\bq)$ for reasons 
connected with the LLL. This formula for $\varepsilon_\bk$ differs {}from
that of other authors, discussed in Sec.\ \ref{recent}, in that it depends
explicitly on the occupation numbers of the other $\bk$ states, and does
not reduce to the self interaction of a dipole even for small
$\bq=\bk-\bk'$ in the integral in the exchange term. Our $\xi_\bk$ gets
its $\bk$ dependence {}from the exchange effect, while the interaction of
the particle with the correlation hole that surrounds it (due to the
vortices) is a ``Hartree-like'' term (and not simple Hartree)
(see \cite{read89} where exchange effects were explicitly neglected). 
Thus the exchange effect found here in the simplest approximation seems to
be complementary to the interaction with the correlation hole, and
probably both terms would be present in a better approximation. As for the
dipolar form of density, we will see that the density does take on this
form, and this could be included in the exchange self energy, but this
would necessitate a complicated self-consistent calculation which could
not be done analytically. In any case, the dipolar effect changes the
form of the interaction at small $\bq$, while intermediate $\bq$
values are important in the exchange self energy. Thus the expression here
is a convenient starting point, and not badly wrong physically, at least
in some cases, as we will see shortly.

The zero-temperature HF dispersion relation can be studied
in detail. Apparently, no difficulties are caused by the absence of a
bare $\varepsilon_\bk$ term. For any repulsive interaction
$\tilde{V}(\bq)=e^{-\frac{1}{2}|\bq|^2}V(\bq)>0$, $\varepsilon_\bk$
increases monotonically with $|\bk|$ for all $\bk$. At $|\bk|=k_F$, 
\bea
\frac{k_F}{m^\ast}\equiv\frac{\partial\xi_\bk}{\partial|\bk|}&=&
-\int_{|\bk'|<k_F}\frac{d^2k}{(2\pi)^2}
\frac{\partial\tilde{V}}{\partial|\bk|}(\bk'-\bk)\non\\
&=&-\frac{k_F}{2\pi}\int
\frac{d\theta_{\bk\bk'}}{2\pi}\tilde{V}(\bk'-\bk)\cos\theta_{\bk\bk'}
\label{effmass}
\eea
(note that $\theta_{\bk\bk'}$ parametrises the
angle between $\bk'$ and $\bk$ which are both on the Fermi surface).
For a $\delta$-function (short-range) potential, $V(\bq)=V({\bf 0})$, 
$1/m^\ast$
is positive and finite.
Thus the system is stable against single-particle excitations. For a
Coulomb interaction, $V(\bq)=2\pi e^2/|\bq|$, there is a logarithmic
singularity at $|\bk|=k_F$:
\be
\frac{\partial\xi_\bk}{\partial|\bk|}\sim-\ln|k-k_F|.
\ee
This is very similar to that for the Coulomb interaction in the three
dimensional electron gas at zero magnetic field treated in HF
approximation. In that case, the divergence is unphysical and is removed
by replacing the bare Coulomb interaction in the exchange term by the
screened one, which leaves a finite effective mass and heat capacity 
$C_V\sim\gamma T\sim m^\ast k_F T$. This conclusion of course depends on
the presence of screening due to the nonzero compressiblity of the
electron gas. In the present problem, the existence of such a
compressiblity is one
of the points we wish to study, so we must return to this later. Note,
however, that replacing the unscreened interaction by the dipolar
interaction also cuts off the divergence in the present problem. As
mentioned already, this will also be left for later discussion. For the
time being, we may consider an interaction of shorter-range (decaying as a
faster power) than Coulomb and the effective mass is then finite within
HF.

The question may be raised of whether a charge-density wave
(CDW) instability could take place due to the absence of a bare kinetic 
energy. However, the constraints $\rho^R(\bq)=0$, though not the same  
as $\int d^2k \,c_{\bk-\frac{1}{2}\bq}^\dagger c_{\bk+\frac{1}{2}\bq}=0$, 
may have a similar effect in maintaining the uniform density of the fluid 
within HF (a CDW in the underlying particles cannot be ruled out at some
filling factors, especially $\nu\ll 1$, but may not be describable within
HF for the fermions). Another possible instability is to pairing as in BCS
theory. This has
been argued by PH \cite{haldpriv}, who found numerically that bosons at
$\nu=1$
tend to form a ground state with high overlap with the Pfaffian state, a
paired state which is presumably incompressible. However, for some
interactions, such pairing may either not occur, or be very weak so that
it occurs only at very low energies, and then the present results for the
``normal'' Fermi-liquid-like state will still apply at higher energies,
temperatures, or wavevectors. For the state of electrons at $\nu=1/2$,
experiment and numerical results both indicate that pairing must be either
extremely weak or absent, so there would seem to be a regime to which the
theory would apply, assuming that it can be extended to $\pt>1$. We return
to the issue of pairing in Sec.\ \ref{allorders}.

\subsection{Constraints}
\label{constraints}

In this subsection we begin a fuller and more systematic analysis which
begins {}from the HF approximation but entails a careful study of the role
of the constraints. In the present subsection, we explain a functional
integral method for handling the constraints exactly. Approximation
methods are discussed beginning in the following subsection,where the
starting point is once again HF. The present subsection could be skipped
on a first reading, but does explain why many statements later
in the paper are restricted to nonzero frequencies.  

The constraint operators $G(\bq)$ obey
\bea
\left[G(\bq),G(\bq')\right]&=&-G(\bq+\bq')2i\sin\frac{1}{2}\bq\wedge\bq',\\
\left[H,G(\bq)\right]&=&0.
\label{firstclass}
\eea
These relations have the property that if all $G(\bq)$ are replaced by
zero throughout, as stipulated by the constraint, then they are still
true. Constraints with this property are termed first-class, while others
are termed second class
\cite{henteit}. Second-class constraints lead to modified commutation
relations given by ``Dirac brackets'' in the constrained subspace, and
are generally more awkward to handle. An example is the constraint of
being in the LLL, applied to one or more charged particles in a magnetic
field, which when imposed in the obvious way is second class, and
consequently the coordinates $x$, $y$ of the particle(s) end up not
commuting when projected into the LLL.
By contrast, systems with only first-class constraints can be viewed as
gauge theories and there are very well-developed methods by which they
can be handled \cite{henteit}. The advantage of the PH approach is that,
while the fields are in the LLL {}from the beginning, the only constraints
involved are first class.

The importance of the first-class property of the constraints is that
$G(\bq)$ form a Lie algebra, SU($N$) or $W_\infty$, and are constants of
the motion, $dG(\bq)/dt=0$ for all $\bq$. Thus, before considering them
as constraints, the $G(\bq)$ can be viewed as generators of a symmetry
algebra of the Hamiltonian. As constants of the motion, the conditions
$G(\bq)=0$, if imposed at the initial time, would hold for all other
times. Our procedure, which is a version of the Faddev-Popov functional
integral method, will differ somewhat {}from this, however. To find
thermodynamic properties and correlation functions, we begin with the
partition function,
\be
Z={\rm Tr}_{G=0}\,e^{-\beta(H-\mu \hat{N})},
\ee 
where the trace is restricted to states satisfying the constraints. This
can be written formally as
\be
Z={\rm Tr}\,e^{-\beta(H-\mu \hat{N})}\delta_{G,0},
\ee
where the trace is taken in the Hilbert space, the Fock space of the
fermions $c$, with no restriction on the fermion number $\hat{N}$. (The
$\mu \hat{N}$ term is included to make this look
conventional, even though the constraints fix $\hat{N}=N$, so the
constrained ensemble is canonical, not grand canonical.) The
$\delta$-function, which imposes all the constraints, can be given a
Fourier representation which essentially, for a nonabelian group, means
integration over the group manifold. Here we return to the U($N$)
notation that we had for finite $N$:
\be
\delta_{G,0}=\int[U^{-1}dU]\,U\int_0^{2\pi}\frac{d\theta}{2\pi}\,
e^{i\theta(\hat{N}-N)},
\ee
where the first integration is over SU($N$) with the invariant (Haar)
normalized measure $[U^{-1}dU]$, and the second is over U(1) and imposes 
$\hat{N}=N$. We can write $U=e^{-\sum_a i\beta\lambda_a G_a}$ (where
$a=1$, \ldots, $N^2-1$ runs over a basis of the SU($N$) Lie algebra)
and convert the unrestricted $\rm Tr$ to a functional integral in the
standard way to obtain
\widetext
\top{-2.8cm}
\be
Z=\int{\cal D}[c,c^\dagger][U^{-1}dU]\beta\frac{d\lambda_0}{2\pi}
\,\exp\left[-\int_0^\beta d\tau\left\{{\rm Tr}\,c^\dagger\frac{d}{d\tau}c
+H-\mu \hat{N}-i\sum_a\lambda_a G_a
-i\lambda_0(\hat{N}-N)\right\}\right],
\label{functintgaugfixd}
\ee
where $H$, $\hat{N}$, and $G_a$ are given by the standard forms in terms
of the Grassman variables $c_{mn}(\tau)$, $c_{nm}^\dagger(\tau)$, and the 
trace in the exponent is on the U($N$) indices. The
commutation properties (\ref{firstclass}) were used in obtaining this
expression. The $\lambda_a$'s and $\lambda_0=\theta/\beta$ now play the
role of time-independent scalar potentials in the sense of gauge theory.
The functional integral results {}from gauge fixing a manifestly
gauge-invariant version,
\be
Z=\int{\cal D}[c,c^\dagger]{\cal D}[\phi]
\,\exp\left[-\int_0^\beta d\tau\left\{-{\rm
Tr}(\frac{d}{d\tau}+i\phi)c^\dagger c
+H-\mu \hat{N}\right\}\right],
\label{ungaugfixd}
\ee
\bottom{-2.7cm}
\narrowtext
\noindent
in which $\phi$ stands for all the $\lambda$'s in $N\times N$ matrix
form, is $\tau$-dependent, and is functionally integrated over the U($N$)
Lie algebra. Under a U($N$) gauge transformation $U$, $\phi\mapsto
U^{-1}\phi U +U^{-1}dU/d\tau$. This reduces to the previous integral
(\ref{functintgaugfixd}) by imposing the condition $d\phi/d\tau=0$ inside
the functional integral (we are neglecting Faddeev-Popov determinants).
This condition is not the same as $\phi=0$
(which is often used instead), which cannot be reached by a gauge
transformation {}from an arbitrary $\phi$, since gauge transformations must
be periodic in $\tau$ with period $\beta$. Thus $\int d\tau\,\phi$ cannot
be gauged away to zero. The holonomy $Pe^{i\int d\tau\,\phi}$ ($P$
denotes that the integral is path ordered), which is an 
element of the group U($N$), remains. This holonomy is the combination
$Ue^{i\theta}$ of the earlier integration variables. 
Under a $\tau$-independent gauge transformation it is not invariant:
\be
Pe^{i\int d\tau\,\phi}\mapsto U^{-1}Pe^{i\int d\tau\,\phi}U,
\ee
and so only the set of eigenvalues of this matrix is gauge invariant.
(Note that there are gauge transformations that permute the eigenvalues.)
The integral in eq.\ (\ref{functintgaugfixd}) is over the holonomy, but
can be further gauge-fixed to leave integration over the eigenvalues
only: 
\bea
\lefteqn{\int[U^{-1}dU]e^{i\int d\tau\,\sum_a\lambda_a
G_a}}\non\\
&\rightarrow&
\frac{1}{N!}\int_0^{2\pi/\beta}\prod_{\alpha=1}^N
\frac{d\lambda_\alpha}{2\pi/\beta}
\prod_{\gamma<\delta}\left|e^{i\beta\lambda_\gamma}
-e^{i\beta\lambda_\delta}\right|^2\non\\
&&\times e^{i\int d\tau\,\sum_\epsilon \lambda_\epsilon
G_{\epsilon\epsilon}},
\eea
with the measure well-known in, for example, random matrix theory (which
here has no connection with the similar-looking LJ factors!).

The reduction of the constraint integrals to only zero-frequency fields
shows that at low temperatures, the integration over these fields is
relatively unimportant, since zero frequency is of zero measure in
integrals over frequency that appear in a diagrammatic treatment, as will
be used in the following. The non-zero frequency part of the constraints
$G(\bq,\omega)=0$ will have to come out automatically without help {}from
an integration over a field that enforces it directly (as in the totally
gauge unfixed version eq.\ (\ref{ungaugfixd})). It will be demonstrated
that this occurs in the next subsection. 

Finally we note that when developing the HF approximation as in Sec.\
\ref{HF} (or when taking the saddle point of the functional integral as
in Appendix \ref{hubb}), the Lagrange multiplier $\bar{\lambda}$ is the
saddle
point value of $i\lambda_0$, so the saddle point value of $\lambda_0$ is
imaginary. This phenomenon is common in such treatments.

\subsection{Conserving approximations}
\label{conserving}

In this subsection we return to the approximate treatment begun in Sec.\
\ref{HF}, consider response functions, and address the question of
whether the constraints are satisfied. The central issue is the use of a
so-called conserving approximation, that is an approximation that
satisfies the relevant Ward identities, which express the symmetry under
U($N$) or $W_\infty$ generated by the constraint operators $G(\bq)$.

The appropriate conserving approximation to use for, say, the
density-density response in a normal Fermi liquid depends on the
approximation used for the one-particle properties, that is, the
conserving property involves consistency of approximations for different
properties. It is well-known that the random-phase approximation (RPA)
corresponds in this sense to the Hartree approximation, and perhaps less
well-known that generalized RPA, also called time-dependent HF,
corresponds to the HF approximation (for discussion of conserving
approximations, see e.g.\ \cite{schrieffer,baymkad}; for the generalized 
HF approximation in a FL, see PN, Ch.\ 5). These are sometimes
stated in terms of $\Phi$-derivability, that is approximations that 
can be derived by making an approximation
once and for all for the free energy $\Phi$ (or for the thermodynamic
potential) in the presence of source fields that couple to the
onservables of interest (such as the density), and then obtaining
response functions in the same approximation by taking
functional derivatives with respect to the sources, guaranteeing the 
same sort of consistency. 

The importance of the conserving approximation depnds on the nature of
the problem. In the example of a normal Fermi liquid, the basic symmetry
is conservation of total particle number, which is not broken by Hartree
or HF. The conserving approximation is then needed to ensure that the
Fermi liquid relations are satisfied, providing detailed relations
among physical quantities. By contrast, in a BCS superconductor, the
simplest approximation (which can be viewed as an extension of HF)
violates conservation of particle number, and the conserving
approximation \cite{schrieffer} not only restores gauge invariance
(number conservation) but also leads to the prediction of a collective 
mode, the Anderson-Bogoliubov mode (which is the Goldstone mode
connected with the spontaneous symmetry breaking in the case of
short-range interactions). Thus the use of a correct approximation has
much greater physical consequences in the latter case.    

Turning now to the present problem, the HF approximation
 of Sec.\ \ref{HF} does not break conservation of total particle number
$\hat{N}$. However, the symmetry generators $G(\bq)=\rho^R(\bq)$ for
$\bq\neq0$ are not conserved by the HF approximation as it stands. The
easiest way to see this is that $G(\bq)$ does not annihilate the HF
ground state, which is just the Fermi sea $|FS\rangle$. Thus this state
does not satisfy the constraints $G(\bq)|FS\rangle=0$ for $\bq\neq0$.
It is also clear that the HF effective Hamiltonian, Eq, (\ref{Keff}),
does not commute with these $G(\bq)$. The solution to this problem will
have to use the conserving approximation appropriate to our HF
approximation. Since there is a conserved quantity for all $\bq$, the
results will be even more striking than in cases such as the BCS theory
where only a global symmetry was broken. We note that the Fermi sea can
be made invariant by projecting to the invarian subspace as in Eq.\
(\ref{projtoRR}). However, such a projection necessitates that further
work be numerical. Analytical work, and thus conceptual understanding,
can be achieved only by persevering with the gauge theory approach.
Rather than give up the Fermi sea trial state and the HF energies
and searching for some other, invariant, starting point, we keep it and
take care of the constraints by the following conserving approximation. 

The conserving approximation will be illustrated here by the calculation
of the $\rho^R$--$\rho^R$, $\rho^R$--$\rho^L$ and $\rho^L$--$\rho^L$
imaginary-time response functions (more precisely, the generalized
susceptibilities), defined in Fourier space by
\bea 
\chi_{ij}(\bq,\omega_n)(2\pi)^2\delta(\bq+\bq')&&\beta\delta_{\omega_n 
+\omega_{n'},0}=\non\\
&&\langle\rho^i(\bq,\omega_n)\rho^j(\bq',\omega_{n'})\rangle,
\eea
in which $i$, $j$ can be $R$ or $L$, $\omega_n$ are the usual
Matsubara frequencies, and it is implicit that the connected part of the
function is taken, thus dropping a $\delta$-function term containing
$\langle \rho^i \rangle$'s. The conserving
approximation that corresponds to HF takes the form of the sum of all
ring and ladder diagrams. The Green's function lines in the diagrams are
the HF Green's functions
\be
{\cal G}(\bk,\omega_\nu)=(i\omega_\nu-\xi_\bk)^{-1}.
\ee
The usual Dyson-equation argument leads to formulas in terms of the
one-interaction irreducible susceptibilies, as discussed in Sec.\
\ref{chern}, defined as those diagrams that do not become disconnected
when one interaction line is cut (note that we disregard the Hartree
self-energy diagrams that are implicitly included in out HF Green's
functions, which means we are treating the diagrams here as skeleton
diagrams; such terms would be absent anyway for a long-range interaction
due to the neutralising background). These formulas, which are
completely general, are (all $\chi$'s have the same arguments $\bq$,
$\omega_n$)
\bea
\chi_{LL}&=&\frac{\chi_{LL}^{\rm irr}}{1+\tilde{V}(\bq)\chi_{LL}^{\rm
irr}},\\ 
\chi_{RL}&=&\frac{\chi_{RL}^{\rm irr}}{1+\tilde{V}(\bq)\chi_{LL}^{\rm
irr}},\\
\chi_{RR}&=&\chi_{RR}^{\rm irr}-\chi_{RL}^{\rm irr}
\frac{\tilde{V}(\bq)}{1+\tilde{V}(\bq)\chi_{LL}^{\rm irr}}\chi_{LR}^{\rm
irr}.
\eea
Note also that $\chi_{LR}(\bq,\omega_n)=\chi_{RL}(-\bq,-\omega_n)$. The
conserving approximation is now the statement that the various $\chi^{\rm
irr}$ are to be calculated (for $\omega\neq0$) as the sum of the ladder
diagrams, with the HF Green's functions. Since $\rho^L$ is the physical
density, $\chi_{LL}^{\rm irr}$ is the one of most physical interest for
long-range $\tilde{V}(\bq)$, such as Coulomb interactions. 

We begin with $\chi_{RR}^{\rm irr}$, so as to show that at $\omega\neq0$,
the fluctuations in the constraints $G(\bq)$ vanish in our approximation. 
The Feynman rule for the interaction can be read off in the standard way
\cite{fetter}; it includes the wavevector-dependent phase factor as well
as $\tilde{V}(\bq)$. Also, there is a phase factor in the $\rho^R$
vertices, as in eq.\ (\ref{rhoRfour}). Note that those in the interaction
arise {}from the phase factors in the physical density $\rho^L$, Eq.\
(\ref{rhoLfour}). In the ladder diagrams for $\chi_{RR}^{\rm irr}$ the
structure of the momenta is such that {\em all the phase factors cancel},
as the industrious reader will verify. Note that this is an exact
statement, and not only valid at small wavevectors, whether internal or
external, so the exponential defining the phase factor was not expanded
in a Taylor series. Consequently, {\em for the ladder diagrams for 
$\chi_{RR}^{\rm irr}$ only}, the ladder series is identical to the
same approximation to the irreducible susceptibility in the usual density
\be
\rho(\bq)=\int\frac{d^2k}{(2\pi)^2}\,c_{\bk-\frac{1}{2}\bq}^\dagger
c_{\bk+\frac{1}{2}\bq},
\ee
in a model with Hamiltonian
\bea
H&=&\frac{1}{2}\int
\frac{d^2k_1\,d^2k_2\,d^2q}{(2\pi)^6}\,\tilde{V}(\bq)\non\\
&&\mbox{}\times c_{\bk_1-\frac{1}{2}\bq}^\dagger
c_{\bk_2+\frac{1}{2}\bq}^\dagger
c_{\bk_2-\frac{1}{2}\bq}c_{\bk_1+\frac{1}{2}\bq}
\eea
with no kinetic-energy term. This could be phrased by saying that
there is the ordinary, Galilean-invariant kinetic energy term with zero
magnetic field, but the mass $m_0$ is infinite. We call this latter model
the zero field, infinite mass (ZFIM) model. Note that the HF approximations 
in the two models also coincide, because the phase factors disappeared
there also. In the ZFIM model, $[\rho(\bq),H]=0$ for all $\bq$, so the
model possesses a gauge symmetry, whether or not we wish to impose a
constraint $\rho=$ constant. In fact, if such a constraint were imposed in
this model, there would be no states that satisfied it at all. The reason
(in classical language) is that in a continuum model, any configuration of
point particles clearly has nonconstant density. In a similar model on 
a lattice, solutions to the constraint exist only if the value of the
particle number required by the constraint at each site is an integer, 
since these are the eigenvalues of the number operator for each site. This
cannot be satisfied if we take the continuum limit (zero lattice spacing)
at fixed average density. In our system representing the LLL, which is in
the continuum, many solutions to the constraint do exist, provided we
choose (similarly to the lattice ZFIM model) the constrained value of the
total number to be the same as the range of the right indices $n$, as we
have done. Therefore, in the ZFIM model, we will consider the gauge
symmetry (or conservation of $\rho(\bq)$), but not require a constraint to
be satisfied. 

Explicitly we can write $\chi_{RR}^{\rm irr}$ (or $\chi^{\rm irr}$ in the
ZFIM model) in terms of the ladder sum, which is the solution to an
integral equation (we define here various quantities to be used
afterwards)
\widetext
\top{-2.8cm}
\bea
\chi_{RR}^{\rm
irr}(\bq,i\omega_\nu)&=&-\frac{1}{\beta}\sum_n\int\frac{d^2k}{(2\pi)^2}
\Lambda(\bk,\bq,i\omega_\nu)
{\cal G}(\bk+\frac{1}{2}\bq,\omega_n+\omega_\nu)
{\cal G}(\bk-\frac{1}{2}\bq,\omega_n)\non\\
&=&-\int \frac{d^2k}{(2\pi)^2}
\Lambda(\bk,\bq,i\omega_\nu)
\frac{f(\xi_{\bk+\frac{1}{2}\bq})-f(\xi_{\bk-\frac{1}{2}\bq})}
{\xi_{\bk+\frac{1}{2}\bq}-\xi_{\bk-\frac{1}{2}\bq}-i\omega_\nu}.
\label{inteqchiRR}
\eea
Here $\Lambda(\bk,\bq,i\omega_\nu)$ is a one-particle--irreducible vertex
function,
\bea
\Lambda(\bk,\bq,i\omega_\nu)&=&
1-\frac{1}{\beta}\sum_n\int\frac{d^2k_1}{(2\pi)^2} 
{\cal G}(\bk_1+\frac{1}{2}\bq,\omega_n+\omega_\nu)
{\cal G}(\bk_1-\frac{1}{2}\bq,\omega_n)
\Gamma(\bk_1,\bk,\bq,i\omega_\nu)\non\\
&=&1-\int \frac{d^2k_1}{(2\pi)^2}
\frac{f(\xi_{\bk_1+\frac{1}{2}\bq})-f(\xi_{\bk_1-\frac{1}{2}\bq})}
{\xi_{\bk_1+\frac{1}{2}\bq}-\xi_{\bk_1-\frac{1}{2}\bq}-i\omega_\nu}
\Gamma(\bk_1,\bk,\bq,i\omega_\nu),
\label{inteqLambda}
\eea
which we have written in terms of the particle-hole  
scattering series (the ladders with external Green's function lines
removed),
\bea
\Gamma(\bk,\bk',\bq,i\omega_\nu)&=&\tilde{V}(\bk'-\bk)
-\frac{1}{\beta}\sum_n\int\frac{d^2k_1}{(2\pi)^2} 
\Gamma(\bk,\bk_1,\bq,i\omega_\nu)
{\cal G}(\bk_1+\frac{1}{2}\bq,\omega_n+\omega_\nu)
{\cal G}(\bk_1-\frac{1}{2}\bq,\omega_n)\tilde{V}(\bk_1-\bk')
\non\\
&=&\tilde{V}(\bk'-\bk)-\int
\frac{d^2k}{(2\pi)^2}\Gamma(\bk,\bk_1,\bq,i\omega_\nu)
\frac{f(\xi_{\bk+\frac{1}{2}\bq})-f(\xi_{\bk-\frac{1}{2}\bq})}
{\xi_{\bk+\frac{1}{2}\bq}-\xi_{\bk-\frac{1}{2}\bq}-i\omega_\nu}
\tilde{V}(\bk_1-\bk').
\label{inteqGamma}
\eea
\bottom{-2.7cm}  
\narrowtext   
\noindent
(Note that, in this approximation, the scattering function depends only
on the difference $\omega_\nu$ of the Matsubara frequencies in the
external fermion lines, and this is why we are able to perform the
frequency sums explicitly.) 

Before analyzing these equations in detail, we pause to point out that
for $\bq$, $\omega=i\omega_\nu$ small and real, they have the form
standard in Fermi liquid theory (see Pines and Nozi\`{e}res
(PN) \cite{pn},
and especially Nozi\`{e}res \cite{noz} for the full, formal treatment),
with the approximation that the $f_{\bk\bk'}$ function on the Fermi
surface is taken to be the lowest-order approximation as already given in
Eq.\ (\ref{fkk'}), for spinless fermions, and this is just the content of
the generalized HF approximation (see PN, Ch.\ 5). The Landau parameters
$F_\ell$ are then given by 
\bea
F_\ell&=&{\cal N}(0)f_\ell,\\
f_\ell&=&\int\frac{d\theta_{\bk\bk'}}{2\pi}f_{\bk\bk'}\cos\ell\theta_{\bk\bk'},
\eea
for $\ell\geq0$, where, as before, 
$\bk\cdot\bk'=k_F^2\cos\theta_{\bk\bk'}$ for $|\bk|=|\bk'|=k_F$. In
particular, we notice that, since the density of states at the Fermi
energy ${\cal N}(0)=m^\ast/2\pi$, and since the bare kinetic energy is
zero, comparison with Eq.\ (\ref{effmass}) yields 
\be
F_1=-1.
\ee
This is a particular case of the relation
\be
m^\ast/m_0=1+F_1
\ee
in ordinary two-dimensional Galilean-invariant Fermi liquids with bare
mass $m_0$. We can view the ZFIM model as such a system but with
$m_0=\infty$, {}from which $F_1=-1$ follows. This is the value that would
usually be interpreted as the borderline of stability of the system;
however, usually this view is taken because the bare mass is finite and
the effective mass vanishes, and the latter causes instability. Here the
effective mass is finite, so the system is not unstable, and moreover is
held right at this point by this symmetry. We take it as implying that
the ladder series must be analyzed with even greater attention than usual
to the limit $\omega\rightarrow0$, $\bq\rightarrow0$, particularly for
the $\ell=1$ angular mode. We also point out a contrast with HLR, where
this formula was invoked, but with the bare (or band) mass $m$ in place of
$m_0$, and was connected with Kohn's theorem and the f-sum rule. There
the interesting limit was $m\rightarrow0$ (to send the cyclotron
mode to infinite frequency), rather than $\infty$. The present discussion
is clearly distinct, though it must be related at some deeper level. 

In Fermi liquid theory, relations like that above are derived through
Ward identities connected with symmetries of the problem, and the
symmetries are global, so the relations are most useful only at small
$\bq$ or $\omega$. Next we will derive a Ward-identity relationship
between $\Lambda$ and the self energy $\Sigma$ within the HF
approximation, in a way more directly connected with the symmetry
generated by the $\rho^R$'s, and valid for all $\omega\neq0$ and $\bq$.

First we express the HF approximation as a pair of self-consistent
equations:
\bea
{\cal
G}(\bk,\omega_n)&=&[i\omega_n-(\Sigma(\bk)-\bar{\lambda}-\mu)]^{-1},\\
\Sigma(\bk)&=&-\frac{1}{\beta}\sum_n\int\frac{d^2k_1}{(2\pi)^2}   
\tilde{V}(\bk-\bk_1){\cal G}(\bk_1,\omega_n)\non\\
&=&-\int\frac{d^2k_1}{(2\pi)^2}\tilde{V}(\bk-\bk_1)f(\xi_{\bk_1}),
\eea
where $\xi_\bk=\Sigma(\bk)-\mu-\bar{\lambda}$ as before
(the direct term has been dropped as it plays no role in the following,
for the one-interaction irreducible functions; it is absent anyway for
the long-range interaction case). Then 
\widetext
\top{-2.8cm}
\bea
\Sigma(\bk+\frac{1}{2}\bq)-\Sigma(\bk-\frac{1}{2}\bq)
   -i\omega_\nu
&=&-i\omega_\nu-\frac{1}{\beta}\sum_n\int\frac{d^2k_1}{(2\pi)^2}
\left(\tilde{V}(\bk+\frac{1}{2}\bq-\bk_1)
-\tilde{V}(\bk-\frac{1}{2}\bq-\bk_1)\right){\cal G}(\bk_1,\omega_n)\non\\
&=&-i\omega_\nu-\frac{1}{\beta}\sum_n\int\frac{d^2k_1}{(2\pi)^2}
\tilde{V}(\bk-\bk_1){\cal G}(\bk_1+\frac{1}{2}\bq,\omega_n+\omega_\nu)\non\\
&&\times\left(\Sigma(\bk_1+\frac{1}{2}\bq)-\Sigma(\bk_1-\frac{1}{2}\bq)
-i\omega_\nu\right){\cal G}(\bk_1-\frac{1}{2}\bq,\omega_n),
\label{eigeneq}
\eea
\bottom{-2.7cm}
\narrowtext
\noindent
after shifting dummy variables. But {}from Eqs.\
(\ref{inteqLambda},\ref{inteqGamma}),
$-i\omega_\nu\Lambda(\bk,\bq,i\omega_\nu)$ obeys the same
inhomogeneous integral equation, the solution of which should be unique,
so we conclude that
\bea
i\omega_\nu\Lambda(\bk,\bq,i\omega_\nu)&=&
i\omega_\nu-\Sigma(\bk+\frac{1}{2}\bq)+\Sigma(\bk-\frac{1}{2}\bq)\non\\
 &=& i\omega_\nu-\xi_{\bk+\frac{1}{2}\bq}+\xi_{\bk-\frac{1}{2}\bq},
\eea
which is the desired Ward identity (compare Ref.\ \cite{schrieffer}). The
left-hand side is the vertex function for
$\partial\rho^R(\bq)/\partial\tau$, which should vanish since
$\rho^R(\bq)$ commutes with the Hamiltonian. It implies that, if
$\Lambda$ is viewed as the scattering amplitude for a fermion scattering
off a potential coupling to $\rho^R$, or for creating or destroying a
particle-hole pair, then the amplitude vanishes if both particles are on
shell, that is if their frequencies $i\omega_n$ satisfy
$i\omega_n=\xi_\bk$. This suggests (following a similar argument in
\cite{read85}, that was inspired by \cite{bergweiss}) that in the
on-shell states (energy eigenstates), if they satisfy the
constraints $G(\bq)=0$, then the latter property is actually preserved in 
the time evolution, in spite of its apparent violation in the HF states. 
This of course is because the calculation we have done is not the naive 
one of looking at the states as noninteracting particles, rather we used
the conserving approximation. It appears that the fermion {\em
excitations} can 
after all be viewed as real physical excitations, satisfying the
constraint conditions on physical states, even though the {\em operators}
$c^\dagger$ are not gauge-invariant and so would connect invariant to
noninvariant states. These physical fermion excitations, which are 
dressed by the fluctuations around the HF states, are the physical 
composite or (as we shall see) neutral fermions discussed in 
\cite{read94} and in Sec.\ \ref{physical}. 

Now we return to our original goal of calculating $\chi_{RR}^{\rm irr}$
in the ladder approximation. Using the Ward identity and Eq.\
(\ref{inteqchiRR}), and assuming $\omega\neq0$, we find
\bea
\chi_{RR}^{\rm
irr}(\bq,i\omega_\nu)&=&\frac{1}{i\omega_\nu}\int\frac{d^2k}{(2\pi)^2}
\left(f(\xi_{\bk+\frac{1}{2}\bq})-f(\xi_{\bk-\frac{1}{2}\bq})\right)\non\\
&=&0.
\eea
Another response function containing $\rho^R$ that should vanish is
$\chi_{RL}^{\rm irr}(\bq,i\omega_\nu)$.
In this case, the appearance of $\rho^L$ in place of one $\rho^R$ implies
that the phase factors do not all cancel, and on using the Ward identity
for the $\rho^R$ vertex we obtain
\bea
\chi_{RL}^{\rm
irr}(\bq,i\omega_\nu)&=&\frac{1}{i\omega_\nu}\int\frac{d^2k}
{(2\pi)^2}
\left(f(\xi_{\bk+\frac{1}{2}\bq})\right.\non\\
&&\left.-f(\xi_{\bk-\frac{1}{2}\bq})\right)
e^{i\bk\wedge\bq}    \non\\
&=&0,
\eea
since shifting $\bk$ by $\mp \frac{1}{2}\bq$ has no effect on the
phase factor.

As promised we have shown that the conserving approximation guarantees
that there are no fluctuations in $\rho^R(\bq)$, at least for nonzero
frequency. For zero frequency, the Lagrange multiplier fields
$\lambda(\bq)$ (or the subset of diagonal elements, according to the
final gauge-fixed form) enter to give the same result, but we will not     
show this explicitly. Similar issues were addressed extensively in the
literature on slave bosons and heavy fermions in the 1980's (see for
example \cite{readnewns83,read85,aublev,milllee,cole,hrw}), and later in
connection with theories of high $T_c$ superconductors and quantum
magnets. These problems also involve constraints, but these are usually
abelian and generate only U(1). It is still frequently stated incorrectly
in the literature that in the functional-integral saddle-point approach to
such problems, ``the constraints are satisfied only on the average''. In
fact, as was well-known to several workers (such as the cited authors) in
the field in the 1980's, the correct RPA or $1/N$ (i.e., conserving)
treatment of fluctuations yields just the same sort of results we have
just derived, namely the vanishing of the vertex function for, and of all 
correlation functions containing, the constraint operators (like our
$G(\bq)$), to all orders in the fluctuations. Thus {\em the average of,
and all fluctuations in, the constraints vanish}, which means that the
constraints are satisfied in every order of approximation, when this is
set up correctly. (The extension to all orders for the present problem
will be discussed later.) 
    
It remains to examine $\chi_{LL}^{\rm irr}$. This will be undertaken in
the next two subsections.

\subsection{Asymptotics of the ladder series}
\label{asymptotics}

In this subsection we continue the analysis of the conserving
approximation of the last subsection. We examine the behavior of the 
ladder series at small $\bq$ and $\omega_\nu$, first to elucidate the
mechanism behind the vanishing of $\chi_{RR}^{\rm irr}$, and then, in the
following subsection, the results are applied to the calculation of the
physical density-density response function $\chi_{LL}^{\rm irr}$.

The equation for $\Gamma$ can be rewritten  
\widetext
\top{-2.8cm}
\be
\int\frac{d^2k_1}{(2\pi)^2}\,\left\{(2\pi)^2\delta(\bk'-\bk_1)+
\tilde{V}(\bk'-\bk_1)
\left(\frac{f(\xi_{\bk_1+\frac{1}{2}\bq})-f(\xi_{\bk_1-\frac{1}{2}\bq})}
{\xi_{\bk_1+\frac{1}{2}\bq}-\xi_{\bk_1-\frac{1}{2}\bq}-i\omega_\nu}\right)
\right\}\Gamma(\bk,\bk_1\bq,i\omega_\nu)=\tilde{V}(\bk-\bk')
\label{fredholm}
\ee
\bottom{-2.7cm}
\narrowtext
\noindent
which shows that it is a Fredholm integral equation, where the integral
kernel appears in the curly brackets on the left-hand side, and
contains $\bq$ and $\omega_\nu$ as parameters. It implies that $\Gamma$
is $\tilde{V}$ times the inverse integral operator. The inverse could be
calculated by finding the eigenvalues and eigenfunctions of the integral
operator on the left.

At $i\omega_\nu=0$, (which could be viewed as the limit
$i\omega_\nu\rightarrow0$), one zero eigenvector can be found for all
$\bq$ by use of the Ward identity proved in the previous subsection; it
is $\xi_{\bk+\frac{1}{2}\bq}-\xi_{\bk-\frac{1}{2}\bq}$ (see Eq.\
(\ref{eigeneq})). Thus for small $i\omega_\nu$, we expect to have, for
all $\bq$, an eigenvector approximately $\xi_{\bk+\frac{1}{2}\bq}
-\xi_{\bk-\frac{1}{2}\bq}$, with eigenvalue tending to zero with
$i\omega_\nu$. If $\bq\rightarrow0$ also, we get 
\be
\xi_{\bk+\frac{1}{2}\bq}-\xi_{\bk-\frac{1}{2}\bq}\simeq\bq\cdot\bv_\bk
\ee
where $\bv_\bk=\nabla_\bk \xi_\bk$. At small $\bq$, the nontrivial
part of the integral kernel becomes
\be
\left.\tilde{V}(\bk'-\bk_1)\frac{\partial f}{\partial
\varepsilon}\right|_{\xi_\bk},
\ee
which for zero temperature $T$ is concentrated at $k=k_F$ (indeed, for
all $\bq$, the difference of Fermi functions is non-zero only in a shell
of width of order $q$ around $k_F$). But this limit of the kernel is
independent of $\bq$, so in addition to the eigenfunction just found
which is proportional to $\cos\theta_\bk$ on the Fermi surface, there is
another proportional to $\sin\theta_\bk$. Note that these eigenfunctions,
in the spirit of a Fermi-liquid analysis in terms of $\delta n_\bk$ or a
deformation of the Fermi surface, are just rigid displacements of the
Fermi sea, respectively parallel and perpendicular to $\bq$. The second
eigenfunction is not a zero mode for $\bq\neq0$, so is expected to
acquire an eigenvalue that is nonzero as $i\omega_\nu\rightarrow0$, but
vanishes as $\bq\rightarrow0$.

For general values of the ratio $i\omega_\nu/q$ the integral equation
and the eigenvalue problem are not easy to analyse, even for
$i\omega_\nu$, $\bq$ small, where the eigenvalue equation takes the form 
\widetext
\top{-2.8cm}
\be
A(\bk,\bq,i\omega_\nu)+\left.\int\frac{d^2k_1}{(2\pi)^2}\,\tilde{V}(\bk-\bk_1)
\frac{\bq\cdot\bv_{\bk_1}}{\bq\cdot\bv_{\bk_1}-i\omega_\nu}\frac{\partial
f}{\partial
\varepsilon}\right|_{\xi_{\bk_1}}A(\bk_1,\bq,i\omega_\nu)
=\lambda(\bq,i\omega_\nu)A(\bk,\bq,i\omega_\nu).
\ee 
\bottom{-2.7cm}
\narrowtext
\noindent
This form of equation is standard in Fermi liquid theory, with
$\tilde{V}(\bk-\bk_1)$ replaced by $-f_{\bk\bk_1}$. At $T=0$, $\partial
f/\partial\varepsilon=-\delta(\xi_\bk)$ and the equation can in principle
be solved for $\bk$ on the Fermi surface, and {\em these values of the
eigenfunction determine it elsewhere}. Accordingly we might expand both
$A$ and $\tilde{V}$ in terms of Fourier modes $\cos\ell\theta_\bk$,
$\sin\ell\theta_\bk$, $\ell=0$, $1$, \ldots, for $|\bk|=k_F$. For
$i\omega_\nu/|\bq|v_F\neq0$, the Fourier modes are mixed by the integral
kernel, so that all components of 
\bea
-\tilde{V}(\bk-\bk')&=&f_0+2\sum_{\ell=1}^\infty f_\ell
\cos\ell\theta_{\bk\bk'}\non\\
&=&f_0+2\sum_{\ell=1}^\infty
f_\ell(\cos\ell\theta_\bk\cos\ell\theta_{\bk'}\non\\
&&+\sin\ell\theta_\bk
\sin\ell\theta_{\bk'})
\label{landpars}
\eea
are involved. We have seen that the $\ell=1$ mode and $f_1$ are crucial
to the analysis and must be kept. The other Landau parameters $F_\ell$
take no special values and merely produce finite renormalizations of the
response functions (some identities are implied by the existence of the
zero mode for all $\bq$, but these bring in derivatives of $\bv_\bk$ and
thus parameters that lie outside of Fermi liquid theory). We propose
just to drop these effects so as to obtain the simplest possible
approximation that is still conserving. This can be done by replacing
$f_\ell$ for $\ell\neq 1$ by zero, or more accurately by assuming that
the only eigenfunctions $A$ that are needed are just
$\bq\cdot\bv_\bk/q$, $\bq\wedge\bv_\bk/q$ (which are the correct
continuations off $|\bk|=k_F$). We will actually use this even to higher
order in $q$, as we will see is necessary.

With this further approximation, the eigenvalues corresponding to the two
eigenfunctions can be evaluated. The final result for $\Gamma$
is
\be
\Gamma(\bk,\bk',\bq,i\omega_\nu)=\frac{\bq\cdot\bv_\bk\,\bq\cdot\bv_{\bk'}}
{\omega_\nu^2\chi_0(\bq,i\omega_\nu)}
-\frac{\bq\wedge\bv_\bk\,\bq\wedge\bv_{\bk'}}
{q^2\chi_0^\perp(\bq,i\omega_\nu)},
\label{gammaaprox}
\ee
where
\be
\chi_0(\bq,i\omega_\nu)=-\int\frac{d^2k}{(2\pi)^2}\,
\frac{f(\xi_{\bk+\frac{1}{2}\bq})-f(\xi_{\bk-\frac{1}{2}\bq})}
{\xi_{\bk+\frac{1}{2}\bq}-\xi_{\bk-\frac{1}{2}\bq}-i\omega_\nu}
\ee
is the ``density-density'' response function of a Fermi gas with
dispersion $\xi_\bk$, and 
\bea
\chi_0^\perp(\bq,i\omega_\nu)&=&-\frac{1}{2}{\cal N}(0)v_F^2
-\int\frac{d^2k}{(2\pi)^2}\,
\left(\frac{\bq\wedge\bv_\bk}{|\bq|}\right)^2\non\\
&&\times\frac{f(\xi_{\bk+\frac{1}{2}\bq})-f(\xi_{\bk-\frac{1}{2}\bq})}
{\xi_{\bk+\frac{1}{2}\bq}-\xi_{\bk-\frac{1}{2}\bq}-i\omega_\nu}
\eea
is the transverse ``current-current'' response function of the same Fermi
gas, including the $\bq$-, $i\omega_\nu$-independent contact
(``diamagnetic'') term. $\chi_0$ arose in a similar way {}from
the longitudinal current-current response, on using the continuity
equation. Note that what we are calling the ``density'' and
``current'', though natural in appearance, are {\em not} to be identified
with the physical density and current. 

The above expressions for $\chi_0$ and $\chi_0^\perp$ are valid for any
$\bq$ and $i\omega_\nu$. On the real frequency axis, at
$\omega/q v_F$ and $\bq$ small, they become
\bea
\chi_0(\bq,\omega+i0^+)&=&{\cal N}(0)+i{\cal N}(0)\omega/(qv_F)\\
\chi_0^\perp(\bq,\omega+i0^+)&=&q^2\chi_{\rm
d}^\ast+i\omega k_F/(2\pi q).
\eea
Here $\chi_{\rm d}^\ast$ is the diamagnetic susceptibility of the Fermi
gas with dispersion $\xi_\bk$. It is a non-Fermi-liquid property that
involves derivatives of $\bv_\bk$ at $k_F$; if $\xi_\bk$ were
$=(\bk^2-k_F^2)/2m^\ast$, then $\chi_{\rm d}^\ast$ would be $=-1/(12\pi
m^\ast)$. These imply that the eigenvalues of the longitudinal and
transverse eigenmodes of the integral kernel above vanish in the ways
predicted in this limit. This involved the cancellation of the
diamagnetic term in the current-current response in both cases; this
cancellation is well-known in normal fluids (i.e., non-superfluids). 

We can now show that even this further approximation is conserving in the
sense discussed in Sec.\ \ref{conserving}. Using the above form of
$\Gamma$, we can calculate 
\be
\chi_{RR}^{\rm irr}=\chi_0-\chi_0(\chi_0)^{-1}\chi_0=0,
\ee 
where the second term is the contribution of $\Gamma$, for all $\bq$ and
$i\omega_\nu\neq0$. In this calculation, the transverse mode in $\Gamma$
did not contribute. A similar calculation shows that $\chi_{RL}^{\rm
irr}=0$. An exact treatment of the ladder series in the regime
$\omega/qv_F\ll 1$ and $q\ll k_F$ yields the same form with
all $\chi_0$'s replaced by $\chi_0/(1+F_0)$, and the cancellation still
occurs, in agreement with the previous subsection.

\subsection{Physical response functions}
\label{physresp}

In this subsection we calculate $\chi_{LL}^{\rm irr}$, the physical
density-density response function, and its limits, the compressibility
and longitudinal conductivity. We also consider the scattering of the
fermions by an external potential, and the expression for the current
density. 

\subsubsection{Density-density response function}
\label{densdensresp}

As already remarked, the fact that the $\rho^L$ vertex contains the
opposite phase factor {}from that in $\rho^R$ means that not all the phase
factors cancel in $\chi_{LL}^{\rm irr}$, instead those at the two vertices
at the ends of the ladder are doubled. We have
\widetext
\top{-2.8cm}
\be
\chi_{LL}^{\rm irr}=\chi_0+\int\frac{d^2k\,d^2k'}{(2\pi)^4}
\frac{f(\xi_{\bk+\frac{1}{2}\bq})-f(\xi_{\bk-\frac{1}{2}\bq})}
{\xi_{\bk+\frac{1}{2}\bq}-\xi_{\bk-\frac{1}{2}\bq}-i\omega_\nu}
\Gamma(\bk,\bk',\bq,i\omega_\nu)  
\frac{f(\xi_{\bk'+\frac{1}{2}\bq})-f(\xi_{\bk'-\frac{1}{2}\bq})}
{\xi_{\bk'+\frac{1}{2}\bq}-\xi_{\bk'-\frac{1}{2}\bq}-i\omega_\nu}
e^{i\bk\wedge\bq-i\bk'\wedge\bq}.
\ee
However, by comparison with $\chi_{RR}^{\rm irr}=\chi_{RL}^{\rm irr}
=\chi_{LR}^{\rm irr}=0$, this simplifies to 
\bea
\chi_{LL}^{\rm irr}&=&-\int\frac{d^2k}{(2\pi)^2}(e^{i\bk\wedge\bq}-1)
(e^{-i\bk\wedge\bq}-1)\frac{f(\xi_{\bk+\frac{1}{2}\bq})-f(\xi_{\bk-\frac{1}{2}
\bq})}{\xi_{\bk+\frac{1}{2}\bq}-\xi_{\bk-\frac{1}{2}\bq}-i\omega_\nu}
+\int\frac{d^2k\,d^2k'}{(2\pi)^4}
(e^{i\bk\wedge\bq}-1)
\frac{f(\xi_{\bk+\frac{1}{2}\bq})-f(\xi_{\bk-\frac{1}{2}\bq})}
{\xi_{\bk+\frac{1}{2}\bq}-\xi_{\bk-\frac{1}{2}\bq}-i\omega_\nu}\non\\
&&\times\Gamma(\bk,\bk',\bq,i\omega_\nu)  
\frac{f(\xi_{\bk'+\frac{1}{2}\bq})-f(\xi_{\bk'-\frac{1}{2}\bq})}
{\xi_{\bk'+\frac{1}{2}\bq}-\xi_{\bk'-\frac{1}{2}\bq}-i\omega_\nu}
(e^{-i\bk'\wedge\bq}-1).
\label{chiirrsimp}
\eea
\bottom{-2.7cm}
\narrowtext
\noindent
For small $\bq$, we now expand the phase factor. The first term is then
the form found by \cite{sm,dhlee,ph}. It is the same as putting
$\rho^L-\rho^R$ in place of $\rho^L$, which goes as $\sim i\bk\wedge\bq$
at
small nonzero $\bq$. The second term is the ladder series with the
insertion $(\bk\wedge\bq)(\bk'\wedge\bq)$ at the two vertices. 
This exhibits the
effectively dipolar nature of the coupling of an external scalar
potential to the physical density: the fermions carry a dipole moment
$\wedge\bk$, as found in Refs. \cite{read94,sm,dhlee,ph} and discussed in
Sec.\ \ref{review}. In $\Gamma$, only the transverse mode now
contributes, and we obtain
\bea
\chi_{LL}^{\rm irr}&=&
q^2m^\ast(\bar{\rho}+m^\ast\chi_0^\perp(\bq,i\omega_\nu))\non\\
&&\mbox{}-\frac{q^2(\bar{\rho}+m^\ast\chi_0^\perp(\bq,i\omega_\nu))^2}
{\chi_0^\perp(\bq,i\omega_\nu)}\non\\
&=&-q^2\bar{\rho}(\bar{\rho}+m^\ast\chi_0^\perp(\bq,i\omega_\nu))/
\chi_0^\perp(\bq,i\omega_\nu).
\label{irrdensresp}
\eea
Note that in the numerator, the $\bar{\rho}$'s occur because of the
absence of a ``diamagnetic'' term to cancel it, and in writing the
remainder of the numerator as $\chi_0^\perp$ we have neglected the
difference between $\bk/m^\ast$ and $\bv_\bk$, which affects the
coefficient of the term in $\chi_0^\perp$ quadratic in $\bq$. This term
can be neglected anyway in the following. 
In the small $\omega/(qv_F)$, $\bq$ region we then have
\be
\chi_{LL}^{\rm irr}(\bq,\omega+i0^+)=\frac{\bar{\rho}^2}{-\chi_{\rm
d}^\ast-i\omega k_F/(2\pi q^3)}.
\label{physdensresult}
\ee
This is similar in form to the result obtained by HLR, or the
renormalized version of it according to the scenario discussed in Sec.\
\ref{chern}, if we note that $\bar{\rho}=1/(2\pi\pt)$ in general (and
$\pt=1$ here), except that the $1$ in the denominator in Eq.\
(\ref{densrespinchern}) has been dropped. That $1$ came {}from the
Chern-Simons term, which couples longitudinal and transverse
fluctuations; by contrast, in the conserving approximation in the 
present approach, the ladder propagator $\Gamma$ does not couple these
modes. 
Note that the first term in the first line of Eq.\
(\ref{irrdensresp}) is essentially the result of Refs.
\cite{sm,dhlee,ph},
\be
\chi_{LL}^{\rm
irr}=q^2m^\ast\left(\bar{\rho}+m^\ast\chi_0^\perp(\bq,i\omega_\nu)\right),
\label{wrongdensresp}
\ee
which behaves differently at low $\omega$ and $\bq$, as we will see.

We now take various limits of this expression. As $\omega\rightarrow0$,
we obtain
\be
\frac{dn}{d\mu}\equiv\lim_{|\bq|\rightarrow0}\chi_{LL}^{\rm irr}(\bq,0)
=-\bar{\rho}^2/\chi_{\rm d}^\ast,
\ee 
which is finite and positive, so the system is compressible as in HLR,
though again the expression differs {}from that in the scenario of Sec.\
\ref{chern}, as given in Eq.\ (\ref{cscomp}). Though
we used the approximate form for $\Gamma$, our result is exact
within the ladder (conserving) approximation.

To obtain the low-frequency longitudinal conductivity, relevant to the
surface acoustic wave experiments, we define a relevant limit:
\be
\sigma_{xx}(\bq)=\lim_{\omega/q\rightarrow0}
\lim_{\stackrel{q\rightarrow0}{\omega/q{\rm\,fixed}}}
\frac{-i\omega}{q^2}\chi_{LL}^{\rm irr}(\bq,\omega+i0^+)
\ee
for $\bq$ parallel to $\hat{x}$ (the conductivity should always be viewed
as the response to the total electric field, so it is related to the
irreducible response). Here ``lim'' means that we keep the leading
nonzero term. 
This limit corresponds to considering a long-wavelength sound wave, so
$|\bq|$ is small $\ll k_F$ and $\omega=|\bq|v_s$, and then taking the 
sound velocity $v_s$ to zero, (i.e.\ $v_s\ll v_F$).
Then we obtain
\be
\sigma_{xx}(q)=\bar{\rho}^2\frac{2\pi q}{k_F}=\frac{q}{2\pi k_F}
\ee
{\em in exact agreement with HLR} for $\pt=1$. There a different procedure
was used to
define $\sigma_{xx}(q)$, as given by HLR eq.\ (B4.a). That and the
present definition give the same result both in the RPA of HLR and in the
present approximation. This result was expected to be very robust on
Fermi liquid grounds, within the scenario discussed in Sec.\ \ref{chern},
since it corresponds to the transverse conductivity of an ordinary
Fermi liquid, which is unrenormalized in Fermi liquid theory. Remarkably,
it is the same here, in spite of other differences in the structure of
the expressions. This result is not obtained {}from the expression
(\ref{wrongdensresp}) \cite{sm}. It is also remarkable how the factor
$\bar{\rho}$, which came {}from a standard gauge-invariance result for the
usual Fermi liquid, here plays one of the roles played in
the CS theory by $\sigma_{xy}$ ($=\bar{\rho}$ in our units). This
effect, that the ``current'' response at $\omega/q\rightarrow0$ of a Fermi
gas to a scalar potential coupled to the dipolar expression for the
density gives the Hall conductivity, was pointed out by St\"{o}rmer
\cite{stormer}.

Finally the spectral density for $\chi_{LL}^{\rm irr}(\bq,\omega)$
implied by Eq.\ (\ref{physdensresult}) is at low frequency
\be
{\chi_{LL}^{\rm irr}}''(\bq,\omega)=\frac{\omega
k_F\bar{\rho}^2/(2\pi q^3)}{{\chi_{\rm
d}^\ast}^2+\omega^2k_F^2/(2\pi q^3)^2}
\ee
(but vanishes for $|\omega|/(qv_F)>1$) and has a peak, an overdamped
mode at $\omega\sim |\bq|^3$, similar to the result of HLR. As many
physicists have noticed, this implies for the various moments, as
$q\rightarrow0$,
\bea
\int_0^\infty{\chi_{LL}^{\rm irr}}''\omega^n&\sim& q^{n+3},\,n\geq
1;\non\\
                                       &\sim&q^3\ln1/q,\,n=0;\non\\
                                       &\sim&{\rm const}\,\,\,n=-1.  
\eea
For $n<-1$, the moments diverge as usual. 

The $n=1$ moment can be obtained exactly, because of the Kramers-Kronig
relation,
\bea
\chi_{LL}^{\rm
irr}(\bq,\omega+i0^+)&=&\int_{-\infty}^{\infty}\frac{d\omega'}{\pi}
\frac{{\chi_{LL}^{\rm irr}}''(\bq,\omega')}{\omega'-(\omega+i0^+)}\non\\
&\sim&\frac{-1}{\omega^2}\int_{-\infty}^{\infty}\frac{d\omega'}{\pi}
\omega'{\chi_{LL}^{\rm irr}}''(\bq,\omega'),
\eea
as $\omega\rightarrow\infty$. The high-frequency behavior of
$\chi_{LL}^{\rm irr}$ at small $q$ can be obtained by returning to the
integral equation for $\Gamma$, Eq.\ (\ref{fredholm}). To leading order in
$qv_F/\omega$, $\Gamma(\bk,\bk',0,\omega)=\tilde{V}(\bk-\bk')$, and we
obtain {}from Eqs.\ (\ref{landpars},\ref{chiirrsimp})
\be
\chi_{LL}^{\rm irr}(\bq,\omega+i0^+)\sim
\frac{-q^4k_F^2\bar{\rho}(1+F_2)}{4\omega^2m^\ast}.
\ee
(The same result except that $F_2$ is replaced by zero is obtained using
our earlier approximation for $\Gamma$.) This can be compared with the
result in a usual Fermi liquid, which is
$-q^2\bar{\rho}(1+F_1)/(\omega^2m^\ast)=-q^2\bar{\rho}/(\omega^2m)$ on
using $1+F_1=m^\ast/m$. We return in Sec.\ \ref{allorders} below to the
question of the general validity of our result, beyond the ladder
approximation.

The moments of the spectral density of the full response function
$\chi_{LL}$ can now be obtained also. For the $n=-1$ moment, one finds
$\sim\tilde{V}(\bq)^{-1}$ for a long range interaction, as usual in a
compressible system. The $n=0$ moment behaves as $q^3\ln1/q$ again, and
gives the LLL ``static'' (equal time) structure factor $\bar{s}(\bq)$. It
does not go as $q^4$ as GMP suggested it should in any liquid state. This
is because compressible liquids have both low-energy modes and long-range
correlations that produce nonanalytic behavior of $\bar{s}(\bq)$. GMP
concluded that fluids in the LLL should be incompressible, but this
argument is invalid (this point has also been made by Haldane
\cite{haldpriv}). The $n=1$ moment goes as $q^4$, as argued by GMP, and
using the high frequency behavior of $\chi_{LL}(\bq,\omega)$, and because
$\tilde{V}(\bq)$ is less singular than $q^{-4}$,
\bea
\int_{-\infty}^{\infty}\frac{d\omega'}{2\pi}
\omega'{\chi_{LL}}''(\bq,\omega')&=&
\int_{-\infty}^{\infty}\frac{d\omega'}{2\pi}
\omega'{\chi_{LL}^{\rm irr}}''(\bq,\omega')\non\\
&=&\frac{q^4k_F^2\bar{\rho}(1+F_2)}{8m^\ast},
\label{f2sumrule}
\eea
to leading order in $q$. GMP found a formula for this moment in terms of
$V(\bq)$ and $\bar{s}(\bq)$, so we obtain a relation among the quantities
$m^\ast$, $F_2$, and $\bar{s}(\bq)$. The result for the $n=1$ moment of
$\chi_{LL}^{\rm irr}$ can also be viewed as a sum rule for the
leading part at small $q$ of the longitudinal conductivity ${\rm
Re\,}\sigma_{xx}(\bq,\omega)=\omega{\chi_{LL}^{\rm
irr}}''(\bq,\omega)/q^2$.

\subsubsection{Fermion scattering vertex}
\label{fermscat}

We now consider the scattering of the fermions by an external potential $V_{\rm
ext}(\br,t)$. The scattering of a fermion {}from wavevector $\bk+
\frac{1}{2}\bq$ to 
$\bk-\frac{1}{2}\bq$ is given in the same ladder diagram approximation by the
vertex function, similar to $\Lambda$ earlier except for a phase factor,
\widetext
\top{-2.8cm}
\be
\Lambda^L(\bk,\bq,i\omega_\nu)=e^{i\bk\wedge\bq}
-\int\frac{d^2k_1}{(2\pi)^2}\,
e^{i\bk_1\wedge\bq}\frac{f(\xi_{\bk_1+\frac{1}{2}\bq})-f(\xi_{\bk_1-
\frac{1}{2}\bq})}
{\xi_{\bk_1+\frac{1}{2}\bq}-\xi_{\bk_1-\frac{1}{2}\bq}-i\omega_\nu}
\Gamma(\bk_1,\bk,\bq,i\omega_\nu) 
\ee
\bottom{-2.7cm}
\narrowtext
\noindent
after removing the same phase on the external lines as for $\Lambda$ (only the
irreducible part is shown). If the phase factors are replaced by 1, we obtain
$\Lambda$, so we will first reconsider this briefly. 

Earlier we showed that, in the small $\bq$ limit, 
\be
\Lambda=1-\bq\cdot\bv_\bk/(i\omega_\nu).
\ee
In terms of the asymptotics of $\Gamma$, the second term is the correction 
produced by the longitudinal mode. While the first term is the bare scalar 
coupling to the external potential, the second term couples to the fermions 
through their velocity, that is to the ``current'' (in the same sense as 
before), and so can be viewed as describing a longitudinal vector potential. 
Because of the factor $\bq/i\omega_\nu$, the vector potential cancels the 
direct effect of the scalar potential, if we consider the electric field they 
produce. The system responds by producing a longitudinal response purely in 
the form of a vector potential, because we chose the gauge such that the 
scalar potential in the functional integral vanishes at nonzero frequencies. 
Thus for gauge-invariant response functions, such as $\chi_{RR}$ that we 
considered earlier, these terms produce complete cancellation, as we saw 
earlier in the example. This should also be true in other calculations, such 
as for the effect of an external ``impurity'' potential on the conductivity, 
if it coupled to $\rho^R$ instead of to $\rho^L$ as it would in fact (such an 
``impurity'' potential would be static, but as usual the same effects would be 
found there as for all nonzero frequencies, thanks to the zero-frequency 
Lagrange multiplier or scalar potential field).    

Since the vertex functions $\Lambda$ and $\Lambda^L$ differ only by the phase
factor, we conclude that the phase factors like $e^{i\bk\wedge\bq}$ can be 
replaced by $e^{i\bk\wedge\bq}-1$ when using $\Lambda^L$. To first order in 
$\bq$, this gives the dipolar coupling $\bk\wedge\bq$ with dipole moment 
$\wedge\bk$. The first term in $\Lambda^L$ is thus the direct coupling of 
$V_{\rm ext}$ to the dipole moment of the fermions. It should be contrasted 
with the direct, minimal coupling to the fermions with charge 1 in the 
scenario for the low energy behavior in the approach of HLR , described in 
Sec.\ \ref{chern}. In the second term in $\Lambda^L$, where the ladder series 
$\Gamma$ contributes, the dipolar coupling brings in the transverse mode in 
the ladder series, as in the calculation of $\chi_{LL}^{\rm irr}$. This 
coupling gives essentially 
\be
\bq\wedge\bv_\bk(\bar{\rho}+m^\ast\chi_0^\perp(\bq,i\omega_\nu))/
\chi_0^\perp(\bq,i\omega_\nu)
\ee
at small $\bq$, $i\omega_\nu$, which is a coupling to the transverse current, 
and is similar to that found in HLR and also in
\cite{kz} in connection with the effects of an impurity potential, that is the
$i\omega_\nu=0$ limit. As there, the external potential couples to the density,
which induces a transverse vector potential, which, because it is singular 
at $\bq=0$, scatters the fermions much more effectively than the direct
minimal coupling to the potential, let alone the dipolar coupling. The
scattering produced can be simplified by comparison with the physical
density $\rho^L$ induced by the same external potential, which is
$\langle\rho^L\rangle-\bar{\rho}=\chi_{LL}^{\rm irr}V_{\rm ext}
(\bq,i\omega_\nu)e^{-\frac{1}{4}|q|^2}$. This shows that if the induced
transverse vector potential is denoted $\ba+\bA$, then we have
\be  
\nabla\wedge\ba=-\langle\rho^L\rangle/\bar{\rho}
=-2\pi\pt\langle\rho^L\rangle,
\label{curlaeqrho}
\ee
which is exactly the equation in the CS theory! This shows that {\em the 
fermions experience a vector potential that obeys eq.\ (\ref{curlaeqrho}),
where $\rho^L$ is the physical charge density,} even though there is no CS
term in the effective gauge field coupling and the fermions behave as
dipoles. This agrees with the use in Refs.\ \cite{read89,read94} of
the Berry phase argument of Arovas {\it et al.} \cite{asw} to obtain the
vector potential seen by the fermions, which in no way assumed that there
are flux tubes attached to the particles, unlike the CS approach. Note
that, since we also have 
\be
\rho^L=\bar{\rho}-\nabla\wedge\bg,
\ee
this is consistent with $\ba+\bA=\bg/\bar{\rho}$ for the longitudinal
part. There should also be an equation $-\dot{\ba}-\nabla
a_0=2\pi\pt\wedge\bj^L$, where ${\bf j}^L$ is the physical current
density. The problem of the form of ${\bf j}^L$
in the present approach will be considered in Subsec.\ \ref{current}. 

\subsubsection{Effect of impurities}
\label{impurities}

Here we consider the effect of impurity scattering on the density-density
response and the longitudinal conducitivity. The HF and ladder
approximations can be reconsidered with impurities present. We neglect
here the mechanism of the preceding subsection, and take only direct
scattering by the impurities, analogously to the bare HF considered so
far. The average self energy should contain an impurity line (the
self-consistent Born approximation), and the ladders contain both impurity
lines and interactions as the rungs of the ladder. The effective mass and
the diamagnetic susceptibility will generally be renormalized by the
impurity effects, but we will not distinguish them {}from their 
counterparts in the pure system. Calculations are straightforward, and the
results can be written down using well-known formulas. The scattering rate
$1/\tau$ is given by the usual expression, but contains $m^\ast$ {}from the
density of states (this could be replaced by the rate {}from the mechanism
of the preceding subsection, but this makes little difference). At $q=0$,
we have
\be
\sigma_{xx}(0,\omega)=\frac{i\omega\bar{\rho}(\bar{\rho}+
m^\ast\chi_0^\perp)}{\chi_0^\perp},
\ee
and, in the Drude approximation, recalling that the current-current
response is isotropic at $q=0$, 
\be
\chi_0^\perp(0,\omega+i0^+)=\frac{i\omega\bar{\rho}\tau}{m^\ast
(1-i\omega\tau)}.
\ee
Then
\be
\sigma_{xx}(0,\omega)=\bar{\rho}m^\ast/\tau=\sigma_0
\ee
independent of $\omega$. This can be viewed as the usual form
of resistivity of the fermions,
$\rho_{xx}=(\bar{\rho}\tau/m^\ast)^{-1}$, divided by $\rho_{xy}^2$, so is
consistent for small $\rho_{xx}$ with the result of the CS theory, of
adding the fermion and CS resistivities (see Eq.\ (\ref{iofflark})). The
frequency-independence is also consistent with this, if in the CS approach
one uses $m^\ast$ in place of $m$, and includes FL corrections as in the
scenario described in Sec.\ \ref{chern}. The effect of the latter
corrections is to replace $1-i\omega\tau$ by $1-i\omega\tau m/m^\ast$ (see
PN, p.\ 191). As $m/m^\ast\rightarrow0$, with $m^\ast$, $\tau$ fixed, the
result above is obtained. 

For finite wavevector, we will consider only the small $\omega$ and $q$
region. With impurities present, $\chi_0^\perp$ is analytic in $q^2$ and
$\omega$, 
\be
\chi_0^\perp(\bq,\omega+i0^+)=q^2\chi_{\rm
d}^\ast+i\omega\bar{\rho}\tau/m^\ast.
\ee
We then obtain the longitudinal conductivity
\be
\sigma_{xx}(\bq,\omega+i0^+)=\frac{i\omega\sigma_0}{i\omega-Dq^2},
\ee
which exhibits a diffusion pole, with diffusion constant
\be
D=-m^\ast\chi_{\rm d}^\ast/(\bar{\rho}\tau),
\ee  
and $\sigma_0$ obeys the Einstein relation $\sigma_0=D\,dn/d\mu$.

\subsubsection{Physical current density}
\label{current}

We turn here to the calculation of the expression for the physical current
density within linear response. The most obvious way to obtain the current
is by projecting the usual expression to the LLL, as was considered by
GMP. This yields
\be
\bj_c=\wedge\nabla\rho^L/(2m),
\ee
which involves the bare mass, and describes the current due to the
cyclotron motion of the particles. Since it clearly obeys
$\nabla\cdot\bj_c=0$, and gives zero when integrated across a section
with a boundary condition of zero density, it does
not contribute to transport. This current, when coupled linearly to a
change in the vector potential, $\bA\cdot\bj_c$, describes a magnetic
moment on each particle, which should be recovered in the U(1) CS
approach, as argued by the authors of Ref.\ \cite{simonhalp}, and
obtained by SM \cite{sm}. 

We are concerned with transport and with response functions,
and this part of the current contains explicit derivatives, so is of less
interest at long wavelengths. We therefore turn to the current due to
drift motion of the guiding centers of the cyclotron orbits of the
particles, due to both the external one-body potential $A_0$, and the
interparticle two-body interaction. We will not consider fully the
response to a change in the physical vector potential $\bA$. The existence
of both parts of the drift current was recognized by GMP and in Ref.\
\onlinecite{read89}; for further discussion, see Refs.\
\cite{stonebook,martinezstone}. In principle, they can be obtained by
carrying the calculation of the projected current to higher order in
$1/\omega_c$ (the cyclotron current $\bj_c$ being the leading term, of
order $\omega_c$), by considering virtual excitation of the particles to
higher Landau levels. This is carried out in Ref.\ \cite{rajsond}; it
yields two types of terms of order $\omega_c^0$ in the matrix elements of
the current within the LLL, for an external potential $V_{\rm ext}$. The
first of these, called $\bj_L^1$, can be
written as a series of derivatives of the LLL-projected potential $V_{\rm
ext}$ and of the density $\rho^L$; the series can be further
divided into a series of exponential form that agrees with the ``Noether
current'' of Martinez and Stone, and another series, beginning with a
third-order derivative, that is of the form of an integral of an
exponential. The second type of term \cite{rajsond} consists of the
modification of the cyclotron current by the effective LLL Hamiltonian to
order $\omega_c^{-1}$, so is more complicated. The general expression for
the current is thus by no means simple. However, to find the net current
for transport purposes, we require only the small $q$ limit, and for this 
the result is just
\be
\bj^L=-\rho^L\wedge\nabla A_0
\ee
for a slowly varying potential $A_0=V_{\rm ext}$, which exhibits the Hall
conductivity $\sigma_{xy}=\bar{\rho}$ in our system. 

For the small $q$ drift current due to the interaction, we have in Fourier
space
\be
\bj^L(\bq)=\int\frac{d^2q'}{(2\pi)^2}i\wedge\bq'\tilde{V}(\bq')
:\rho^L(\bq+\bq')\rho^L(-\bq'):
\label{exactcurrent}
\ee
Diagrammatically, one can see that to calculate the linear response
current to a scalar perturbation within the conserving approximation, 
it will be sufficient to take the operator itself in the HF approximation. 
Since $\langle\bj^L\rangle=0$ in the unperturbed ground state, the leading
term is obtained by replacing a pair of operators $c^\dagger$, $c$ by
their expectation value in the ground state,
\be
\langle c^\dagger_{\bk_1} c_{\bk_2}\rangle =(2\pi)^2\delta(\bk_1-\bk_2)
\theta(k_F-k_1),
\ee
in all possible ways, that is two ``direct'' and two ``exchange'' terms.
Of the direct terms, one vanishes and the other is seen to give the Hall
current produced by the field due to the interaction with the average
density of particles at wavevector $\bq$,
\be
\bj^L(\bq)_{\rm
direct}=-i\wedge\bq\,\bar{\rho}\,\tilde{V}(\bq)\rho^L(\bq).
\ee
In calculating the irreducible response to the total field, this term is
clearly included automatically. Therefore we can turn to the exchange
terms which alone give the irreducible response. Since $q$ is small, we
use
\bea
\theta(k_F-|\bk+\frac{1}{2}\bq|)&-&\theta(k_F-|\bk
-\frac{1}{2}\bq|)\non\\
&=&-q\cos\theta_\bk\delta(k-k_F)
\eea
for $\bq$ in the $\hat{\bf x}$ direction, and after some algebra we obtain
\bea
\bj^L(\bq)_{\rm irr}&=&\int\frac{d^2kd^2k'}{(2\pi)^4}i\wedge(\bk-\bk')
\tilde{V}(\bk-\bk')c^\dagger_{\bk-\frac{1}{2}\bq}c_{\bk+\frac{1}{2}\bq}\non\\
&&\times(-q\cos\theta_{\bk'}\delta(k'-k_F))\non\\
&=&-\int\frac{d^2k}{(2\pi)^2}\left[i\wedge\bk\,\bq\cdot\bk(1+F_2)/m^\ast\right.
    \non\\
&&\left.\mbox{}+i\wedge\bq\,k_F^2(F_0-F_2)/(2m^\ast)\right]
c^\dagger_{\bk-\frac{1}{2}\bq}c_{\bk+\frac{1}{2}\bq},\non\\
&&
\eea
where the Landau parameters $F_\ell$ were defined earlier, in Eq.\
(\ref{landpars}). We assumed that only values of $\bk$ near $k_F$ will be
used, which is true for linear response (thus $k_F^2=\bk^2$). 

Interpreting $c^\dagger_{\bk-\frac{1}{2}\bq}c_{\bk+\frac{1}{2}\bq}$ as $\delta
n_\bk(\bq)$ in FL theory, where $\delta n_\bk(\br)$ is the departure of
the distribution of occupied $k$ values at $\br$ {}from the ground state,
and is assumed to be nonzero only for $\bk$ near $k_F$, this can be
identified as
\be
j_\mu^L(\bq)_{\rm
irr}=-i\varepsilon_{\mu\nu}q_\lambda\Pi_{\nu\lambda}(\bq)
\label{currentform}
\ee
where
\bea
\Pi_{\mu\nu}&=&\int\frac{d^2k}{(2\pi)^2}\left[(k_\mu k_\nu
-\frac{1}{2}k^2\delta_{\mu\nu})(1+F_2)/m^\ast\right.\non\\
&&\left.\mbox{}+\frac{1}{2m^\ast}k^2\delta_{\mu\nu}(1+F_0)\right]\delta
n_\bk(\bq)
\label{stress}
\eea
is the stress or momentum flux tensor of the FL; it is equivalent to that
in Ref.\ \cite{LL10}, modified to two dimensions. Since we have
identified $\rho^L(\br)=\bar{\rho}-\nabla\cdot{\bf P}$ and ${\bf
P}(\br)=\wedge\bg(\br)$, we expect a term in the current $\bj^L_{\rm
irr}=\dot{\bf P}(\br)$ \cite{dhlee}. But by momentum conservation, 
\be
\frac{\partial g_\mu}{\partial t}+\partial_\nu\Pi_{\mu\nu}=0,
\ee
and so we find Eq.\ (\ref{currentform}). Since we also wish to identify
$\ba+\bA=\bg/\bar{\rho}$, we find
\be
\bj^L_{\rm irr}=\bar{\rho}\wedge\dot{\ba},
\ee
which is essentially the other CS-like equation. 

We should also add to the Hamiltonian the potential terms
\be
\int\frac{d^2q}{(2\pi)^2}\left[a_0(\bq)\rho^R(-\bq)+\tilde{A}_0\rho^L(-\bq)
\right]
\ee
where $a_0$ is the scalar potential introduced earlier, which implements
the constraint $\rho^R=\bar{\rho}$, and for which we chose the gauge
$\dot{a}_0=0$, and $\tilde{A}_0(\bq)=e^{-\frac{1}{4}q^2}A_0(\bq)
=\tilde{V}_{\rm ext}(\bq)$ is the externally applied potential. Then 
the right hand side of the momentum conservation equation becomes 
\be
-(\bar{\rho}\nabla a_0+\rho^L(\br)\nabla A_0)
\ee
at long wavelengths. Here the coefficient $\bar{\rho}$ arises {}from
$\rho^R$ on using the constraint. There is also a similar Hall
contribution to $-\rho^L\wedge\nabla A_0$ to the current density $\bj^L$.
Expressing the total physical current $\bj^L$ in terms of
$\dot{\bg}=\dot{\ba}/\bar{\rho}$, we obtain
\be
\bj^L=\bar{\rho}\wedge(\dot{\ba}+\nabla a_0),
\ee
which is manifestly gauge-invariant and of the CS form.

If we consider the current in the right coordinates, $\bj^R$, in a similar
way, we find that in the absence of $a_0$ it vanishes identically, because
$\rho^R$ commutes with $H$. This result of vanishing current was already
invoked in Sec.\ \ref{overview}. It can be interpreted by breaking the
current into the pieces $\bg/m^\ast$ and $(\ba+\bA)\bar{\rho}/m^\ast$
shown there. The first term represents the velocity of the fermions, while
the second represents the usual backflow correction in a FL, which in the
present case of $F_1=-1$ exactly cancels the first part. The same effect
occurs in the ZFIM model: the total current carried by each fermion is
$\bk/m_0$ by Galilean invariance, and $m_0=\infty$, so it vanishes.
(In the presence of $a_0$, we find $\bj^R=\bar{\rho}\wedge\nabla a_0$, a
Hall current. This does not affect our argument in Sec.\ \ref{overview},
which uses only the irreducible part of the current, {}from interactions.) 
A similar calculation can be given for $\bj^L$. The velocity term and the
leading part of the backflow are the same as for $\bj^R$, and so cancel.
The subleading terms then give the result as calculated above. This
cancellation of the leading terms is (perhaps not surprisingly) similar to
what occurred in the formula for the density $\rho^L$ on using the
constraint on $\rho^R$. 

The irreducible longitudinal current density-density response function 
$\chi_{j^L_x\rho^L}^{\rm irr}$ should be $\omega/q$ times 
$\chi_{LL}^{\rm irr}$. This can be verified in terms of the ladder series
expressions for both, if one consistently either keeps or drops the Landau
parameters $F_\ell$ for $\ell\neq 1$ in both the ladder series and the  
$\bq\cdot\bj^L$ vertex. In particular, in the small $q/\omega$ limit, the
$(1+F_2)/m^\ast$ term in $\bj^L$ reproduces that in $\chi_{LL}^{\rm irr}$. 
However, if we consider the longitudinal current-current response, which
should be $\omega^2/q^2$ times $\chi_{LL}^{\rm irr}$, we see that the
two-point current correlation function starts at higher order in
$q/\omega$ than the required term (two-point correlation functions always
vanish as $\omega\rightarrow\infty$). A similar difficulty is familiar in
the usual Fermi liquid and is resolved by the presence of a term in the
current,
$-\bar{\rho}\bA/m$ (the ``diamagnetic current''), that is  linear in the
applied vector potential perturbation, so that the response function
($\chi_0^\perp$ in the noninteracting case) consists of a constant plus
the two-point function of the current without the $\bA$ term. A similar
effect should occur here. The term required in $\bj^L$ is of order $q^2$.
One might attempt to find such possible terms by making the stress tensor
expression Eq.\ (\ref{stress}) gauge invariant by replacing all
$\bk$'s (including $k_F^2=\bk^2$) by $\bk-\ba-\bA$. This does not affect
the other calculations done up to now because, in the absence of a
perturbation in the external $\bA$, the net $\ba+\bA$ does not contribute
in linear response. But further work is required to check the form of this
tensor, since the gauge invariance under SU($N$) or $W_\infty$ reduces to
conventional U(1) gauge theory only at long wavelengths, while this
expression for $\bj^L$ is higher order in derivatives. In any case, such
minimal coupling terms do not produce the necessary factors of $q$, and
so should be absent. A way
to find the part of the {\em longitudinal} current linear in a change in
$\bA$, which should be correct at long wavelengths, is to add a term 
$-\delta\bA\cdot\bj^{L(0)}$ (where $\bj^{L(0)}$ is the exact expression 
(\ref{exactcurrent}) of zeroth order in the perturbation $\delta\bA$) to
the Hamiltonian, then calculate the longitudinal current through first
order terms in $\delta\bA$ by commuting $\rho^L$ with $H$. The resulting
first order term can be seen to give the correct high-frequency limit of
the response, because it is given by a double commutator of $H$ with
$\rho^L(\pm\bq)$, which is what appears in the sum rule for the first
moment of the spectral density of $\chi_{LL}^{\rm irr}$, and we have seen 
that it is related to $(1+F_2)/m^\ast$ also. Thus the correct
term is obtained, and must be used in the longitudinal
current-current response for all $\omega/q$ to ensure agreement with the
density-density response. 

We now consider the full conductivity tensor at $\bq=0$. The longitudinal
part has already been considered. The full conductivity tensor can be
written in the Kubo form
\be
\sigma_{\mu\nu}(0,\omega+i0^+)=\bar{\rho}\varepsilon_{\mu\nu}
+\frac{1}{i(\omega+i0^+)}\chi_{j_\mu^L j_\nu^L}^{\rm irr}(0,\omega+i0^+)
\ee
where the first term is the Hall conductivity and $\chi_{j_\mu^L 
j_\nu^L}^{\rm irr}$ is the current-current two-point function for the
irreducible part of the current. This form was proposed by Lee
\cite{dhlee}. We may also consider the conductivity tensor when
impurities are present. Note that the $q^2$ term in $\bj^L$ does not
contribute when $\bq=0$, even when impurities are present. However, we
expect an additional contribution to $\bj^L$ {}from the impurity potential, 
which we have not explicitly calculated. Because averaging (using Gaussian
disorder) produces diagrams like those for interactions, except that no
frequency is transferred along impurity lines, it should be similar to
that derived above. It will represent the loss of conservation of momentum
when disorder is present. Only the off-diagonal part of $\chi_{j_\mu^L
j_\nu^L}^{\rm irr}$, or the corresponding transverse response to a scalar
perturbation, has not so far been calculated. Because the
ladder diagrams in the interaction and impurity lines do not violate
parity (reflection symmetry), there can be no off-diagonal terms unless
the impurity current vertices that we have not calculated contain pieces
both parallel and perpendicular to $\bq$. If such terms are absent, then 
$\sigma_{xy}=\bar{\rho}$, unaffected by impurities in this approximation. 
As emphasized by Lee \cite{dhlee},
this differs {}from the result of the U(1) CS approach mentioned in Sec.\ 
\ref{chern}. It was argued in Ref.\ \cite{krotov} that in the U(1) CS
fermion approach, applied to the $\nu=1/2$ case, particle-hole symmetry 
implies that $\sigma_{xy}=1/2$ exactly, which is only satisfied by the
scenario described in Sec.\ \ref{chern} if $\sigma_{\psi xy}$ of the
CS fermions is $-1/2$. Assuming our results also apply to $\nu=1/2$, there
is clearly no problem with particle-hole symmetry in our self-consistent
Born approximation (SCBA). We should point out, however, that in this or
the similar approximation for the U(1) CS approach, the results do agree
at leading order in $\rho_{\psi xx}/\rho_{xy}$, and the condition
$\sigma_{\psi xy}=-1/2$ is only needed to guarantee $\sigma_{xy}=1/2$ to
{\em all} orders in this expansion. Thus the contrast between the naive
SCBA result $\sigma_{\psi xy}=0$ and the required $\sigma_{\psi xy}=-1/2$
is not such a dramatic singular correction as it might appear at first
sight. At higher orders there will of course be other correction terms not
included in the SCBA, which can drive the system into the critical regime
representing the transition between quantized Hall plateaus.

\section{Extension to all orders in the interaction, and discussion}
\label{allorders}

In this Section, we consider the extension of the results of Sec.\
\ref{HFconserving} to all orders in the interaction, and describe the
structure of the results we expect, in a scenario which replaces the
previous U(1) CS scenario described in Sec.\ \ref{chern}. First we
consider a more complicated conserving self-consistent approximation, with
special attention to long-range interactions. Then we explain the FL
theory structure for sufficiently long-range interactions. 

In the HF and generalized HF approximation of Sec.\ \ref{HFconserving},
the exchange diagrams contained the bare interaction $\tilde{V}(\bq)$,
and this led to a vanishing $m^\ast$ at $k_F$ for Coulomb or longer-range
interactions. An obvious improvement to make is to insert the ladder
series into the Coulomb vertex, as in Sec.\ \ref{fermscat}. The
longitudinal part of the ladder series $\Gamma$ renders the coupling to
the fermions dipolar at long wavelengths, which removes the divergence in
$1/m^\ast$ for interactions less singular than $1/q^3$. At the same time,
we can insert the ladder series inside the interaction line itself, thus
screening the interaction. We can also {\em replace} the interaction line
in the exchange diagram by $\Gamma$. Finally, we make this approximation 
self-consistent by making these replacements for {\em all} interaction
lines, including those in $\Gamma$, thus iterating to self-consistency.
This approximation, applied to response functions as well as the self energy, 
is once again conserving in the same sense as in Sec.\ \ref{conserving},
and the conclusions there, which follow {}from $F_1=-1$, still apply. 

This approximation is clearly not as tractable as HF, but we can still
make some general statements. The system should still be compressible for
all interactions considered (those less singular than $1/q^2$ as
$q\rightarrow0$). The longitudinal mode in the ladder just produces the
dipolar coupling effects already mentioned, which do not cause a breakdown
of FL theory, though the effect of the exchange self energy that contains
$\Gamma$ in place of $V$ has not been calculated. The transverse mode in
$\Gamma$ produces singularities in the self energy for Coulomb or
shorter-range interactions. The self-consistent summation proposed here is
the same as regards the transverse mode as that studied in Refs.\
\cite{hlr,polchinski} (and similar to that in Ref.\ \cite{holstein}). 
We have nothing to add here to the previous discussion of this case,
except to emphasize that these singular effects should be treated {\em
after} the other FL renormalizations discussed in this paper, and that in
relation to the U(1) CS approach, the effects incorporated in this paper
are related to the longitudinal, not transverse, CS gauge field
fluctuations (see Sec.\ \ref{chern}). For interactions longer range than
Coulomb, there is no breakdown of FL theory, since $m^\ast$ remains finite
and the quasiparticle decay rate vanishes faster than the renormalized
excitation energy $\xi_\bk$ as $k\rightarrow k_F$, though not as fast as
$k^2$. 

We can now discuss the general structure expected in the results to all
orders in the interaction; some of this is implicit in the foregoing
discussion. We consider only interactions longer-range than Coulomb, so
there is no breakdown of FL theory. For Coulomb interaction, the results
are probably still useful, since the only other effect is a logarithmic
divergence in $m^\ast$, which is very weak.

For such interactions, we again separate in the response functions the
``direct'' or reducible diagrams, which represent the long-range
self-consistent field produced by the expectation value of the density.
The remaining diagrams are analyzed in terms of the fermion-hole
irreducible scattering vertex, which at $\bq$, $\omega\rightarrow0$ is
nonsingular and defines the parameters $f_\ell$ and hence $F_\ell=m^\ast
f_\ell/2\pi$. A Ward identity, now valid to all orders, implies that
$F_1=-1$. In fact an identity for the $\rho^R$ vertex, like that in Sec.\
\ref{conserving}, is valid to all orders and for all $\bq$ and
$\omega_\nu\neq0$, and expresses the fact that
$[\rho^R,H]=0$. In the general diagrams that contribute to these vertex
functions, the phase factors in the interaction vertex do not all cancel,
so the system is not equivalent to the ZFIM model. The results nonetheless
have the same structure as in Sec.\ \ref{HFconserving}, and at long
wavelengths can be interpreted in terms of an infinitely-strongly coupled
gauge field, coupled to the FL. {\em There are no parity violating effects
in the long-wavelength dynamics of this system}, because the Landau
interaction $f_{\bk\bk'}$ is even under exchange of $\bk$ and $\bk'$. The
only parity-violating effects come in the coupling to external
electromagnetic fields, where the Hall effect appears, and the
physical density and current obey the CS-like equations. The
self-consistent field produced by the long-range interaction (the
reducible terms) also produces Hall currents, but there is no parity
violation because interactions within the system couple to the density at
both ends. The fluctuations in the longitudinal part of the gauge field
can be reconsidered by changing to the gauge $\nabla\cdot\ba=0$, in which
it is the scalar potential $a_0$ that fluctuates (at all frequencies).
This absorbs the $F_0$ we had previously, and the condition
$\rho^R=\bar{\rho}$ is maintained through an effective $F_0$ that is now
infinite (the Landau parametrization is not gauge invariant). The
longitudinal part of the ladder series at low $\omega/q$ gives an
effective interaction between the fermions, which is of order the
inverse density of states (this is similar to effects in the local Fermi
liquid in the Kondo problem, see \cite{readnewns83}). Because the
leading ``monopolar'' part of the $\rho^L$ density fluctuations is
suppressed by this, the leading nontrivial part is described by the subleading,
dipolar part of the exact density expression $\rho^L$ (note that this
subleading coupling is not described by the minimally-coupled
long-wavelength Hamiltonian Eq. (\ref{ourHeff})). A noteworthy feature of
our approach is that this is not obtained separately {}from the transverse
gauge field effects, nor inserted at the beginning, but emerges later. The
dipole moment $\wedge\bk$ on each fermion is not renormalized, because the
momentum is a conserved quantity. This really deserves an explicit proof,
but it will be omitted because of the similarity to results in standard FL
theory (see, e.g., Nozi\`{e}res \cite{noz}); quite generally, conserved
quantities are not renormalized. 

The compressibility is given by 
\be
\frac{dn}{d\mu}=-\frac{\bar{\rho}^2}{\chi_{\rm d}^{\ast}},
\ee 
where $\chi_{\rm d}^\ast$ is the fully renormalized (irreducible)
diamagnetic susceptibility, and is the only non-Fermi surface quantity to
make an appearance in the response in the regime $\bq$, $\omega$ small.
The other quantities mentioned in Sec.\ \ref{HFconserving} are given by
the same forms as there, when written in terms of $\bar{\rho}$, $k_F$,
$m^\ast$, $F_\ell$, and $\chi_{\rm d}^\ast$. In particular, we mention the
longitudinal conductivity in the regime $q^3<\omega<qv_F$, relevant to
surface acoustic waves. The result, which is identical to that of HLR, is
exact in the same way, and for the same reason, as the
low-frequency transverse conductivity of the usual FL. Also, the high
frequency behavior, or $n=1$ moment of the spectral density, of the
irreducible density-density response, is given by the same sum-rule-like
form as in Sec.\ \ref{densdensresp}, as long as we consider only
excitation of a single quasiparticle-quasihole pair (in the FL sense). 
If multiple quasiparticle-hole pairs do not contribute at this order in
$q$, then this ``sum rule'' is exact. In the usual FL, multiple
quasiparticle-hole pairs contribute to spectral densities at $O(q^4)$, by 
considerations of phase space, and
the f-sum rule is for the $q^2$ part (and higher-order terms actually
vanish in this particular case). Thus it is not certain in our case that
our sum rule is exact. The same phase-space considerations apply, and if
we assume that the squared matrix element of the density $\rho^L$ is of
order $q^2$ (i.e., dipolar) for matrix elements to multiple
quasiparticle-hole excitations, as we have seen it is for single
quasiparticle-quasihole excitations, then these other contributions can be
neglected. This seems likely to be correct, but as we do not have a proof,
we will leave it as a conjecture that {\em Eq.\ (\ref{f2sumrule}) is an
exact relation}, which we call the ``$F_2$ sum rule'', and that it holds
for both the irreducible and reducible responses, as in generalized HF.
If correct, we also obtain a relation of $(1+F_2)/m^\ast$ to the LLL
structure factor $\bar{s}(\bq)$ and $\tilde{V}$, as noted already in Sec.\
\ref{densdensresp}. 

When impurities are included, an improved approximation is obtained by
treating them diagrammatically similarly to the interaction lines as
described at the beginning of this Section. In this Drude- (or SCBA-)-like
approximation, the conductivity takes the same form as in Sec.\
\ref{impurities}. Based on the existing results \cite{sm,dhlee,ph}, we
also expect that similar results hold for $\pt>1$, with
$\bar{\rho}=(2\pi\pt)^{-1}$.

We expect that the direct interaction of the particle with its correlation
hole (or attached vortices), described in Refs.\
\cite{read89,read94,sm,dhlee,ph} is contained in this description, but may
not be easily obtainable diagrammatically. If it is obtained in some
approximation, the effects stemming {}from $F_1=-1$ will still be present
when the approximation is conserving. 

One other way that the FL picture could break down is by a pairing
instability as in the theory of superconductivity. The interaction in the
quasiparticle-quasiparticle channel with quasiparticles of wavevectors
$\bk$, $-\bk$ can be considered using the ladder approximation. The
dipolar nature of the coupling gives rise to an attractive interaction,
as noticed by the authors of Refs.\ \cite{ph,haldpriv}. Since the system
is compressible, this interaction is screened. In addition, the ladder
series $\Gamma$, representing transverse and longitudinal gauge field
fluctuations, can be exchanged between the fermions, and the transverse
part can be combined with the interaction $V$. The transverse gauge field
is believed to be pair-breaking when included in an Eliashberg equation
treatment \cite{bonesteel}. The longitudinal part gives an extra repulsive
short-range interaction, which also suppresses pairing, especially in the 
$s$-wave channel. Therefore the
question of whether pairing is actually expected to occur requires careful
consideration. There is unpublished evidence that it does occur for bosons
at $\nu=1$ for some interactions \cite{ph,haldpriv}. If pairing does
occur, the system will become incompressible at low energies and
long wavelengths, essentially because of the
Meissner effect in the superfluid Fermi system: the diamagnetic
susceptibility now behaves as $\chi_{\rm d}^\ast\sim-1/q^2$, which
inserted in our result for $d n/d\mu$ shows the system is incompressible.
This shows that it is not just the symmetries of the Hamiltonian that make
the ground state compressible in the FL-like state, but it is the fact
that the state is assumed to be a normal (non-superfluid) liquid.

Assuming the system is a FL, the scenario we have described here and in
Sec.\ \ref{overview} is essentially a FL coupled to an infinitely-strongly
coupled gauge field (that represent $F_1=-1$), with no CS term. The
central point was the Ward identity that gave $F_1=-1$. We connected this
with the gauge invariance under U($N$)$_R$, or equivalently with
conservation of $G(\bq)$. Other authors have very recently commented on
``translational invariance in momentum space''
\cite{sm,haldpriv,halpsterncomm,dhlee}, and its relation to some sort of
gauge
symmetry. We will try to make this more precise. The Hamiltonian Eq.\
(\ref{phHam}) is invariant under shifts of the wavevectors of all the 
fermions by $\bf Q$: $\bk\rightarrow\bk+{\bf Q}$. The
generator of a translation of the wavevectors of all fermions in such a
system (or in ours) is
\be
\frac{1}{2}i\int\frac{d^2k}{(2\pi)^2}\left[c^\dagger_\bk\nabla_\bk c_\bk
-(\nabla_\bk c^\dagger_\bk)c_\bk\right].
\ee
In first quantization and in position space, it is simply $\sum_i \br_i$.
This is related to Galilean invariance in ordinary systems with finite
bare mass $m_0$. If we rescale the generator of Galilean transformation
\cite{gottfried} to obtain shifts in $\bk_i$ instead of in ${\bf
v}_i=\bk_i/m_0$, we obtain 
\be
\sum_i(\br_i-t{\bf p}_i/m_0),
\ee
and the second term can be dropped when $m_0\rightarrow\infty$. However,
in this limit we obtain the ZFIM model, and the Galilean symmetry is
enlarged to the local gauge symmetry generated by $\rho(\bq)$, already
discussed. In our system, by contrast, the gauge symmetry is generated by 
$\rho^R(\bq)$,
\bea\rho^R(\bq)&=&\int\frac{d^2k}{(2\pi)^2}e^{-\frac{1}{2}i\bk\wedge\bq}
c^\dagger_{\bk-\frac{1}{2}\bq}c_{\bk+\frac{1}{2}\bq}\non\\
&=&\hat{N}+\int\int\frac{d^2k}{(2\pi)^2}\frac{1}{2}\bq\cdot[i\wedge\bk
c^\dagger_\bk c_\bk\non\\
&&\mbox{}+c^\dagger_\bk \nabla_\bk c_\bk-(\nabla_\bk c^\dagger_\bk)
c_\bk],
\eea
keeping {\em all} terms to linear order in $\bq$. Using the similar
expansion of $\rho^L(\bq)$, the generator of shifts in $\bk$ can be
written as the first-order term in $\rho^R+\rho^L$. The other, unused, 
pieces are the particle number $\hat{N}$ and the momentum $\int \bk
c^\dagger_\bk c_\bk$, which are also conserved quantities (note that the
terms in $\rho^L$, $\rho^R$ linear in $\bq$ are generators of magnetic
translations in the left, right coordinates, respectively, written in
momentum space). Thus the ``shifting'' symmetry is part of the gauge
symmetry, {\em in combination with other global symmetries}, and not just
part of the gauge symmetry as stated by SM. Even so, for some purposes,
viewing it just as part of the gauge symmetry can be useful, as we saw in
Sec.\ \ref{overview}, and will again in the next paragraph. 

In unpublished work \cite{haldpriv}, Haldane proposed to write the
effective Hamiltonian of the quasiparticles, for the case of a finite
system on a torus (say a square torus of side $L$), as
\be
H_{\rm eff}=\frac{1}{4m^\ast N}\sum_{ij}(\bk_i-\bk_j)^2,
\label{Hhald}
\ee
which possesses the shifting symmetry. In this system, shifting all the
momenta by the smallest possible amount $2\pi/L$ changes the total
momentum by $2\pi N/L$, and gives a state equivalent to the original one
\cite{hald85}. The latter fact is assumed in numerical calculations, and
such calculations seem to confirm this form of the Hamiltonian. We may
identify this Hamiltonian as similar to our
\be
\sum_i(\bk_i-\ba-\bA)^2/(2m^\ast),
\ee
in the case of a spatially constant $\ba+\bA$, since (by an equation of
motion) $\ba+\bA=\bg/\bar{\rho}=\sum_i\bk_i/N$. The shift transformation
is a gauge transformation (up to caveats just discussed) that does not
change the physical states; this fact goes beyond the simple symmetry
property possessed by (\ref{Hhald}). Our Hamiltonian is preferable
because, when $\ba$ is allowed to vary spatially, it represents a local
interaction, unlike Eq.\ (\ref{Hhald}). Integrating out $\ba$, and using 
the constraint on the density $\rho^R$, we obtain 
a Hamiltonian like that in SM, except that we have the effective mass
$m^\ast$, whereas in their work it appears at a stage where they instead
have the bare mass $m$. This Hamiltonian is also the starting point for
the arguments of Ref.\ \cite{halpsterncomm}.

\section{Conclusion}
\label{conclusion}

In this paper we have developed a truly lowest-Landau-level theory for the
Fermi-liquid-like state of charged bosons at $\nu=1$. We used a formalism
of Pasquier and Haldane \cite{phvar}, in which the composite fermion
fields depend on two complex coordinates, one of which is the coordinate
of the boson, and the other is in effect the coordinate of a vortex in the
wavefunction of the other bosons, attached to the boson. The wavefunctions
in both these coordinates are restricted to the lowest Landau level, and
there are operator constraints which fix the density in the vortex
coordinates. The constraints imply that the system is a gauge theory. The
effective theory for low-energy, long-wavelength phenomena is a Fermi
liquid in which the fermions couple to a gauge field, for which there are
no bare terms in the action. The ladder series treatment in Sec.\
\ref{HFconserving}, with the approximate form Eq.\ (\ref{gammaaprox}), is
equivalent to the RPA applied to this gauge field. Since there is no
Chern-Simons term in the gauge field action, the longitudinal and
transverse modes decouple. The longitudinal part, within RPA, gives rise
to an effective scalar interaction at small momentum exchange of order 
the inverse density of states. This enforces the fixed-density constraint.
The transverse part couples to the physical density, the first nontrivial
term in which is dipolar in form and parity-violating. Each fermion
carries a dipole moment equal to its wavevector. The result is a
finite compressibility, and a low-frequency longitudinal conductivity that
agrees with that in HLR. The gauge field obeys the same Chern-Simons
equations relating it to the physical density and current as in the
U(1) Chern-Simons fermion approach of HLR. Because there is no CS term in
the action, the results nonetheless differ in form {}from those in the
scenario for the fully-renormalized theory based on HLR. Although the
gauge theory reduces to an ordinary U(1) theory at long wavelengths, this
has to be supplemented by the expression for the density, which is a
non-minimal coupling {}from the U(1) point of view. The form of the
expression for the physical current intimates that this is not the whole
story, and we expect that the full $W_\infty$ gauge group will be involved
in general. In view of existing results of other authors
\cite{sm,dhlee}, the results obtained here for $\pt=1$ (bosons at $\nu=1$)
are expected to apply also for other cases of the FL-like state, when
written in terms of $\bar{\rho}=(2\pi\pt)^{-1}$ and other parameters.
There are many possible extensions and applications of the present
methods, to which we hope to return elsewhere.

\acknowledgements
We thank D.~Green, S.~Kivelson, D.-H.~Lee, G.~Murthy, V.~Pasquier,
M.~Stone, and H.~St\"{o}rmer for interesting discussions, and especially
R.~Shankar and D.~Haldane for detailed explanations of their work,
B.~Halperin for prescient hints, and R.~Shankar for many other
discussions. We also thank D.~Green and I.A.~Gruzberg for technical
assistance. This work was supported by NSF grant DMR-9157484.

\appendix

\section{Noncommutative geometry for pedestrians}
\label{noncom}

In this appendix we explain the formalism we use for states in and
operators acting in the Hilbert space of a single particle in the lowest
Landau level, in the simplest case of the infinite plane with uniform
magnetic field, and magnetic length equal to 1 (see also Ref.\ \cite{gj}).
This is equivalent to the ``noncommutative plane'' in noncommutative 
geometry. In particular we explain the ``noncommutative Fourier 
transform'' which we use extensively.

The normalized basis states in coordinate representation in the symmetric
gauge are
\be
u_m(z)=\frac{z^m e^{-\frac{1}{4}|z|^2}}{\sqrt{2\pi 2^m m!}}.
\ee
A general state in the Hilbert space thus has wavefunction 
$\psi(z)=f(z)e^{-\frac{1}{4}|z|^2}$, where $f$ is a complex analytic
function that does not grow too fast at infinity, so that $\int|\psi|^2$
is finite. All operators can be written as integral kernels, so that an
operator $\hat{a}$ is represented by the kernel $a(z,\zb')$, which acts on
states $\psi(z)$ as
\be
\hat{a}\psi(z)=\int d^2 z' a(z,\zb')\psi(\zb'),
\ee
and matrix products become the ``star product'' $\hat{a}\ast\hat{b}$, 
the integral kernel of which is
\be
\hat{a}\ast \hat{b}(z,\zb')=\int d^2z_1\, a(z,\zb_1)b(z_1,\zb').
\ee
The operators themselves can, of course, be expanded as 
\be
a(z,\zb')=\sum_{m,n=0}^\infty a_{mn} u_m(z) \overline{u_n(z')},
\ee
so that $a_{mn}$ are elements of infinite matrices.

Arbitrary operators in the larger Hilbert space of states in all Landau
levels, that is all square-integrable complex functions in the plane
(really, sections of the appropriate bundle), can be projected to the LLL.
In particular, the identity $\delta(\br-\br')$ has matrix elements
$\delta_{mn}$ in the orthonormal basis, and the corresponding operator as
an integral kernel is
\bea
\delta(z,\zb')&\equiv&\sum_m u_m(z) \overline{u_m(z')}\non\\
&=&\frac{1}{2\pi}\exp(-\frac{1}{4}|z|^2-\frac{1}{4}|z'|^2+\frac{1}{2}z\zb').
\eea
As befits the identity, this obeys $\hat{\delta}\psi=\psi$,
$\hat{\delta}\ast\hat{a}=\hat{a}\ast\hat{\delta}=\hat{a}$. This operator
also implements projection to the LLL.

Another operator is defined by multiplication by the plane wave
$e^{i\bk\cdot\br}$. Its projection to the LLL is
\bea
\lefteqn{\int d^2z_1 \delta(z,\zb_1) e^{i\bk\cdot\br_1}
\delta(z_1,\zb')}\non\\
&=&\delta(z,\zb')e^{\frac{1}{2}i(\bar{k}z+k\zb')-\frac{1}{2}|k|^2},
\eea
where, in this appendix, $k=k_x+ik_y$ (elsewhere in the paper $k=|\bk|$
for all vectors $\bk$). It is convenient to define 
\be
\tau_\bk(z,\zb')=\delta(z,\zb')e^{\frac{1}{2}i(\bar{k}z+k\zb')
-\frac{1}{4}|k|^2}.
\ee
Thus $\hat{\tau}_\bk=e^{i\bk\cdot\hat{\bf R}}$, the
adjoint of which is $\hat{\tau}_{-\bk}$, so 
$\overline{\tau_\bk(z',\zb)}=\tau_{-\bk}(z,\zb')$.
The operator $\tau_\bk$ has the effect of magnetic translation (i.e.,
translation which commutes with the Landau level index) by $-ik$ or
$\wedge\bk$ in the plane \cite{girvpriv}. It obeys the well-known 
magnetic-translation relation:
\be
\hat{\tau}_\bk\ast\hat{\tau}_{\bk'}=\hat{\tau}_{\bk+\bk'}e^{\frac{1}{4}
(\bar{k}k'-k\bar{k}')}.
\label{magtrans}
\ee
Here $\frac{1}{4}(\bar{k}k'-k\bar{k}')=\frac{1}{2}i{\rm
Im}\bar{k}k'=\frac{1}{2}i\bk\wedge\bk'$, which is $i$ times the
(signed) area of the triangle formed by $\bk$, $\bk'$, $-(\bk+\bk')$.

The $\tau_\bk$ are the natural functions for use in defining a
``noncommutative Fourier transform''. The motivation is that functions
(like the operator kernels) of $z$ and $z'$ are like wavefunctions for a
single particle in zero magnetic field, for which plane waves make sense.
For such a function $a(z,\zb')$, we write
\be
a(z,\zb')=\int \frac{d^2k}{2\pi} a_\bk \tau_\bk (z,\zb'),
\ee
and for the inverse transformation
\be
a_\bk=\int\, \hat{a}\ast\hat{\tau}_{-\bk},
\ee
where the integral is defined by $\int \hat{b}={\rm Tr}\,\hat{b}=\int
d^2z\,b(z,\zb)$. The
inversion theorem for this transform is easily proved by Gaussian
integration. We note the orthonormality and completeness relations,
\bea
\int\,\hat{\tau}_\bk\ast\hat{\tau}_{\bk'}&=&2\pi\delta(\bk+\bk'),\\
\int \frac{d^2k}{2\pi} \tau_\bk(z,\zb')\tau_{-\bk}(w,\wb')
&=&\delta(z,\wb')\delta(w,\zb').
\eea
The ``noncommutativity'' of the transform shows up when one has
convolutions, where the relation (\ref{magtrans}) must be used.

In the main text the above formalism is applied to second quantized
operators $c$, $c^\dagger$, $\rho^L$, $\rho^R$, where it concerns their
dependence on the $z$, $w$ variables, and has nothing to do with the Fock
space in which they act as operators. In the case studied in this paper,
the Fourier transform can be applied to $c$ and $c^\dagger$ because the
net magnetic field strength vanishes for $\nu=1/\pt=1$. (For $\nu\neq 1$,
one would require the full set of Landau-level states in the net, 
effective magnetic field \cite{jain}, projected to the $z$, $w$
variables, in place of the plane waves which project to $\tau_\bk$.
The Fourier transform would still apply to $\rho^L$ and $\rho^R$, of
course.) For $\nu=1$ we define
\bea
c(z,\wb)&=&\int \frac{d^2k}{(2\pi)^{3/2}} c_\bk \tau_\bk(z,\wb),\\
c_\bk&=&(2\pi)^{1/2}\int\,\hat{c}\ast\hat{\tau}_{-\bk};
\eea
the normalization has been chosen so as to obtain the conventional
anticommutators in Eq.\ (\ref{cananti}). For $\rho^L$ and $\rho^R$ we use
the normalization given above for an arbitrary $\hat{a}$, and the
properties of the $\tau_\bk$'s lead to Eqs.\ (\ref{rhoLfour}),
(\ref{rhoRfour}). We also note that for the diagonal values $z=z'$,
\bea 
\rho^L(z,\zb)=\int\frac{d^2q}{(2\pi)^2} \rho^L(\bq)
e^{i\bq\cdot\br-\frac{1}{4}|q|^2},\non\\
\rho^L(\bq)=e^{\frac{1}{4}|q|^2}\int d^2r \rho^L(z,\zb)e^{-i\bq\cdot\br},
\non
\eea
and similarly for $\rho^R$. This exhibits the connection with GMP.

Finally we note that other formulas of noncommutative geometry can be
obtained in the integral kernel formalism. For example, the
commutator in the star product,
\be
\hat{a}\ast\hat{b}-\hat{b}\ast\hat{a}
=\left[\hat{a}\stackrel{\textstyle\ast}{,}\hat{b}\right], 
\ee
defines the ``Weyl-Moyal bracket'' that generalizes the Poisson bracket of
functions on the classical phase space to the quantum case. It is usually
written as an infinite series of derivatives. Our integral kernel
formulation avoids such series and allows generalization to other (e.g.\
compact) Riemann surfaces, or to nonuniform field strengths. In all cases,
one can begin with an orthonormal set of LLL states, i.e.\ holomorphic
sections of the appropriate bundle. A crucial operator is the
``reproducing kernel'' analogous to $\delta(z,\zb')$. This can be easily
obtained for the sphere and the torus, for uniform field strength. 

\section{Hubbard-Stratonovich transformation and the $1/M$ expansion}
\label{hubb}

Here we show how to reproduce the results of the HF and ladder
approximations as the saddle-point and Gaussian fluctuations in a
Hubbard-Stratonovich field. First, one may replace the interaction term in
the imaginary-time action by
\bea
\int\prod_{i=1}^4
d^2z_i\,[c^\dagger(z_1,\zb_2)c(z_3,\zb_4)V(\br_2-\br_3)\sigma
(z_4,\zb_3,z_2,\zb_1)&&\non\\
\mbox{}+\frac{1}{2}|\sigma(z_4,\zb_3,z_2,\zb_1)|^2V(\br_2-\br_3)],\,\,\,&&
\label{hstrans}
\eea
(the $\tau$-dependence and $\tau$-integration is implicit)
where $\sigma$ is a fourth-rank tensor field, written in the coordinate
notation using LLL orthonormal functions as for $c$, $c^\dagger$, and is
hermitian:
\be
\overline{\sigma(z_4,\zb_3,z_2,\zb_1)}=\sigma(z_1,\zb_2,z_3,\zb_4),
\ee
and integrate functionally over $\sigma$. Performing the latter functional
integral reproduces the interaction term. The field $\sigma$ decouples the
interaction in the exchange channel. The saddle point approximation for
the $\sigma$ integral (along with the Lagrange multipliers) reproduces the 
exchange, but not the Hartree, part of Hartee-Fock. Gaussian fluctuations
in $\sigma$ around the saddle point reproduce the ladder series. Thus the
ladder series becomes the RPA in the $\sigma$ field. It should be possible
to identify part of the $\sigma$ fluctuations as the gauge field, in a
manner similar to that in some lattice models \cite{rs}.

In other problems, such a saddle point and Gaussian fluctuations are the
leading terms in a $1/M$ expansion, where $M$ is the number of components
of a field corresponding to our $c$, $c^\dagger$. We may introduce such
components here, and then set $M=1$ at the end, by replacing $c_{mn}$ by
$c_{mn\alpha}$, where $\alpha=1$, \ldots, $M$. The interaction is taken
independent of $\alpha$, so the system has SU($M$) symmetry. Then Eq.\
(\ref{hstrans}) now has the form
\be
\int\sum_{\alpha=1}^M\,c^\dagger_\alpha c_\alpha V \sigma
+\frac{1}{2}M\int |\sigma|^2 V
\ee
schematically. This appears suitable for $1/M$ expansion, but there is a
problem with the constraints. The latter must still be taken to be 
\be
\sum_{\alpha=1}^M\sum_m c_{nm\alpha}^\dagger c_{mn'\alpha}=\delta_{nn'}
\ee
in order to reproduce an $M$-component system of bosons, whatever the
filling factor.  To obtain zero net field for the fermions, we must be at
total filling factor $\nu=1$, so we must have $\bar{\rho}=1/2\pi$, that is
of order $M^0$, not $M$. Therefore not all the terms in the action are of
order $M$, and we can expect problems with the $1/M$ expansion. These are
not necessarily completely fatal, however; an expansion can sometimes be
obtained even in such cases (see Ref.\ \cite{readnewns83}). It is not
possible to rescale or redefine the model to avoid this problem. It could
be avoided if we could attach $1/M$ of a vortex to each particle (which
would now be anyons, so that $c^\dagger$ still creates fermions), as in
the U(1) CS approach \cite{ioffe}. However, this is not possible in the
present PH formalism.

\widetext



\begin{references}
\narrowtext
\bibitem{book} 
  For a review, see, {\it e.g.}, {\em The Quantum Hall Effect}, 
  edited by R.E.~Prange and S.M.~Girvin (Second Edition, Springer-Verlag, 
  New York, 1990). 
\bibitem{girvin}S.M.~Girvin in Ref.\ \cite{book}.
\bibitem{gm}S.M.~Girvin and A.H.~MacDonald, \prl {\bf 58}, 1252 (1987).
\bibitem{laughan}R.B.~Laughlin, \prl {\bf 60}, 2677 (1988).
\bibitem{zhk}S.C.~Zhang, T.H.~Hansson, and S.~Kivelson, \prl {\bf 62}, 82
    (1989).
\bibitem{jain}J.K.~Jain, \prl {\bf 63}, 199 (1989); \prb {\bf 40}, 8079
    (1989); {\it ibid.} {\bf 41}, 7653 (1990).
\bibitem{lopfrad}A.~Lopez and E.~Fradkin, \prb {\bf 44}, 5246 (1991).
\bibitem{fishlee}D.-H.~Lee and M.P.A.~Fisher, \prl {\bf 63}, 903 (1989).
\bibitem{read87}N.~Read, Bull. Am. Phys. Soc, {\bf 32}, 923 (1987).
\bibitem{read89}N.~Read, \prl {\bf 62}, 86 (1989).
\bibitem{hlr}B.I.~Halperin, P.A.~Lee, and N.~Read, \prb {\bf 47}, 7312
   (1993).
\bibitem{read94}N.~Read, Semicond. Sci. Technol. {\bf 9}, 1859 (1994)
   [cond-mat/9501090].
\bibitem{wilczek}F.~Wilczek, \prl {\bf 49}, 957 (1982).
\bibitem{laugh}R.B.~Laughlin, \prl {\bf 50}, 1395 (1983).
\bibitem{hald83}F.D.M.~Haldane, \prl {\bf 51}, 605 (1983).
\bibitem{halp84}B.I.~Halperin, \prl {\bf 52}, 1583 (1984).
\bibitem{klz}S.~Kivelson, D.-H.~Lee, S.-C.~Zhang, \prb {\bf 46}, 2223
     (1992).
\bibitem{willett90}R.L.~Willett {\it et al.}, \prl {\bf 54}, 112 (1990).
\bibitem{willett93a}R.L.~Willett {\it et al.}, \prb {\bf 47}, 7344 (1993). 
\bibitem{willett93b}R.L.~Willett {\it et al.}, \prl {\bf 71}, 3846 (1993).
\bibitem{stormerferm}W.~Kang {\it et al.}, \prl {\bf 71}, 3850 (1993).
\bibitem{goldman}V.~Goldman, B.~Su, and J.K.~Jain, \prl {\bf 72}, 2065
       (1994).
\bibitem{asw}D.~Arovas, J.R.~Schrieffer, and F.~Wilczek, \prl {\bf 53},
     722 (1984).
\bibitem{sm}R.~Shankar and G.~Murthy, \prl {\bf 79}, 4437 (1997);
    cond-mat/9802244.
\bibitem{dhlee}D.-H.~Lee, cond-mat/9709233.
\bibitem{ph}V.~Pasquier and F.D.M.~Haldane, cond-mat/9712169.
\bibitem{haldpriv}F.D.M.~Haldane, private communication.
\bibitem{mr}G.~Moore and N.~Read, Nucl. Phys. B {\bf 360}, 362 (1991).
\bibitem{rr96}N.~Read and E.~Rezayi, \prb {\bf 54}, 16864 (1996).
\bibitem{il}L.~Ioffe and A.~Larkin, \prb {\bf 39}, 8988 (1989).
\bibitem{kz}V.~Kalmeyer and S.-C.~Zhang, \prb {\bf 46}, 9889 (1992).
\bibitem{sh}S.~Simon and B.I.~Halperin, \prb {\bf 48}, 17368 (1993).
\bibitem{kane}C.L.~Kane, S.~Kivelson, D.-H.~Lee, and S.-C.~Zhang, \prb
  {\bf 43}, 3255 (1991); S.-C.~Zhang, Int. J. Mod. Phys. {\bf B6}, 25
  (1992).
\bibitem{mil}M.~Milovanovic and N.~Read, (unpublished); M.~Milovanovic,
   Ph.D. thesis, Yale University, 1996 (unpublished).
\bibitem{jr}J.K.~Jain and N.~Read, \prb {\bf 40}, 2723 (1989).
\bibitem{zhang}S.-C.~Zhang, Int. J. Mod. Phys. {\bf B6}, 25 (1992).
\bibitem{holstein}T.~Holstein, R.~Norton, and P.~Pincus, \prb {\bf 8},
     2649 (1973).
\bibitem{reizer}M.~Reizer, \prb {\bf 39}, 1602 (1989).
\bibitem{polchinski}J. Polchinski, Nucl. Phys. B {\bf 422}, 617 (1994).
\bibitem{naywil}C.~Nayak and F.~Wilczek, Nucl. Phys. B {\bf 417}, 359
   (1994); {\it ibid.} {\bf 430}, 534 (1994).
\bibitem{khveshchenko}D.V.~Khveshchenko and P.C.E.~Stamp, \prb {\bf 49},
     5227 (1994); D.V.~Khveshchenko, {\it ibid.} {\bf 49}, 16893 (1994).
\bibitem{ganwan}J.~Gan and E.~Wong, \prl {\bf 71}, 4226 (1994).
\bibitem{ioffe}L.B.~Ioffe, D.~Lidsky, and B.L.~Altshuler, \prl {\bf 73},
    472 (1994); B.L.~Altshuler, L.B.~Ioffe, and A.J.~Millis, \prb 
    {\bf 50}, 14048 (1994).
\bibitem{palee}Y.B.~Kim, A.~Furusaki, X.-G.~Wen, and P.A.~Lee, \prb 
    {\bf 50}, 17917 (1994); Y.B.~Kim, P.A.~Lee, and X.-G.~Wen, {\it ibid.}
     {\bf 52}, 17275 (1995). 
\bibitem{sternhalp}A.D.~Stern and B.I.~Halperin, \prb {\bf 52}, 5890
     (1995).
\bibitem{houghton}A.~Houghton, H.-J.~Kwon, and J.B.~Marston, \prb 
      {\bf 52}, 8002 (1995).
\bibitem{read95}N.~Read, Surface Sci. {\bf 361/362}, 7 (1996).
\bibitem{rr94}E.~Rezayi and N.~Read, \prl {\bf 72}, 900 (1994); {\it
    ibid.} {\bf 73}, 1052 (1994).
\bibitem{gj}S.M.~Girvin and T.~Jach, \prb {\bf 29}, 5617 (1984).
\bibitem{gmp}S.M.~Girvin, A.H.~MacDonald and P.~Platzman, \prb {\bf 33},
    2481 (1986).
\bibitem{halpsterncomm}B.I.~Halperin and A.D.~Stern, cond-mat/9802061.
\bibitem{phvar}The treatment here, which is an alternative to that in
   Ref.\ \cite{ph}, was explained to us by Haldane \cite{haldpriv}.
\bibitem{rs}N.~Read and S.~Sachdev, Nucl. Phys. B {\bf 316}, 609 (1989);
  \prb {\bf 42}, 4568 (1990).
\bibitem{fairlie}D.B.~Fairlie, P.~Fletcher, and C.K.~Zachos, Phys. Lett. B
  {\bf 218}, 203 (1989); J. Math. Phys. {\bf 31}, 1088 (1990); D.B.~Fairlie
     and C.K.~Zachos, Phys. Lett. B {\bf 224}, 101 (1989).
\bibitem{winf}For a review of the algebra, see e.g.\ I.I.~Kogan,
  hep-th/9401093; for some applications to the quantum Hall
  effect see
  S.~Iso, D.~Karabali, and B.~Sakita, Phys. Lett. B {\bf 296},
  143 (1992); B.~Sakita, Phys. Lett. B {\bf 315}, 124 (1993); A.~Cappelli,
  C.A.~Trugenberger, and G.R.~Zemba, Nucl. Phys. B {\bf 396}, 465 (1993);
  Phys. Lett. B {\bf 306}, 100 (1993); A.~Cappelli, G.V. Dunne,
  C.A.~Trugenberger, and G.R.~Zemba, Nucl. Phys. B {\bf 398}, 531 (1993).
\bibitem{fetter}A.L.~Fetter and J.D.~Walecka, {\em Quantum Theory of
  Many-Particle Systems} (McGraw-Hill, New York, NY, 1971).
\bibitem{henteit}See, e.g., M.~Henneaux and C.~Teitelboim, {\em
   Quantization of Gauge Systems} (Princeton University Press, Princeton,
   NJ, 1992). 
\bibitem{schrieffer}J.R.~Schrieffer, {\em Theory of Superconductivity}
   (Addison-Wesley, Reading, MA, 1964), p.\ 227--233.
\bibitem{baymkad}G.~Baym and L.P.~Kadanoff, Phys. Rev. {\bf 124}, 287
    (1961).
\bibitem{pn}D.~Pines and P.~Nozi\`{e}res, {\em The Theory of Quantum
   Liquids, Vol.\ 1: Normal Fermi Liquids} (Benjamin, New York, NY, 1966).
\bibitem{noz}P.~Nozi\`{e}res, {\em Theory of Interacting Fermi Systems}
   (Addison-Wesley, Reading, MA, 1964).
\bibitem{read85}N.~Read, J. Phys. C {\bf 18}, 2651 (1985).
\bibitem{bergweiss}B.~Berg and P.~Weisz, Nucl. Phys. B {\bf 146}, 205
     (1978).
\bibitem{readnewns83}N.~Read and D.M.~Newns, J. Phys. C {\bf 16}, 3273
    (1983).
\bibitem{aublev}A.~Auerbach and K.~Levin, \prb {\bf 34}, 3524 (1986).
\bibitem{milllee}A.J.~Millis and P.A.~Lee, \prb {\bf 35}, 3394 (1987).
\bibitem{cole}P.~Coleman, \prb {\bf 35}, 5072 (1987).
\bibitem{hrw}A.~Houghton, N.~Read and H.~Won, \prb {\bf 35}, 5123 (1987).
\bibitem{stormer} H.~St\"{o}rmer, private communication.
\bibitem{simonhalp}S.H.~Simon, A.~Stern and B.I.~Halperin, \prb {\bf 54},
    R11114 (1996).
\bibitem{stonebook} {\em Quantum Hall Effect}, edited by M.~Stone (World
     Scientific, Singapore, 1992).
\bibitem{martinezstone}J.~Martinez and M.~Stone, Int. J. Mod. Phys. {\bf
     7}, 4389 (1993).
\bibitem{rajsond}R.~Rajaraman and S.~Sondhi, Mod. Phys. Lett. B {\bf 8},
    1065 (1994).
\bibitem{LL10}E.M.~Lifshitz and L.P.~Pitaevskii, {\em Physical Kinetics}
     (Pergamon Press, Oxford, 1981), p.\ 315.
\bibitem{krotov}D.-H.~Lee, Y.~Krotov, S.-C.~Zhang, and S.A.~Kivelson, \prb
    {\bf 55}, 15552 (1997).
\bibitem{bonesteel}N. Bonesteel, private communication.
\bibitem{gottfried}K.~Gottfried, {\em Quantum Mechanics, Vol.\ 1:
     Fundamentals} (Benjamin/Cummings, Reading, MA, 1966), p.\ 252.
\bibitem{hald85}F.D.M.~Haldane, \prl {\bf 55}, 2095 (1985).
\bibitem{girvpriv}S.M.~Girvin, private communication.


\widetext
\end{references}
\end{document}